\documentclass[pdflatex,sn-basic]{sn-jnl}

\usepackage{graphicx}%
\usepackage{multirow}%
\usepackage{amsmath,amssymb,amsfonts}%
\usepackage{amsthm}%
\usepackage{mathrsfs}%
\usepackage[title]{appendix}%
\usepackage{xcolor}%
\usepackage{textcomp}%
\usepackage{manyfoot}%
\usepackage{booktabs}%
\usepackage{subfig}%
\usepackage{algorithm}%
\usepackage{centernot}%
\usepackage{algorithmicx}%
\usepackage{algpseudocode}%
\usepackage{listings}%
\usepackage{mathtools}

\theoremstyle{thmstyleone}%
\theoremstyle{thmstyletwo}%

\theoremstyle{thmstylethree}%

\begin{document}

\title[Article Title]{An Asymptotic Analysis of Bivalent Monoclonal Antibody-Antigen Binding}


\author*[1]{\fnm{Luke A.} \sur{Heirene}}\email{luke.heirene@maths.ox.ac.uk}

\author[1]{\fnm{Helen M.} \sur{Byrne}}

\author[2]{\fnm{James W. T.} \sur{Yates}}

\author[1]{\fnm{Eamonn A.} \sur{Gaffney}}

\affil[1]{\orgdiv{Mathematical Institute}, \orgname{University of Oxford}, \orgaddress{\street{Andrew Wiles Building, Radcliffe Observatory Quarter, Woodstock Road}, \city{Oxford}, \postcode{OX2 6GG},  \country{United Kingdom}}}

\affil[2]{\orgdiv{DMPK Modelling, DMPK, Preclinical Sciences}, \orgname{GSK}, \orgaddress{\street{Gunnels Wood Road}, \city{Stevenage}, \postcode{SG1 2NY}, \country{UK}}}


\abstract{Ligand-receptor interactions are fundamental to many biological processes. For example in antibody-based immunotherapies, the dynamics of an antibody binding with its target antigen directly influence the potency and efficacy of monoclonal antibody (mAb) therapies. In this paper, we present an asymptotic analysis of an ordinary differential equation (ODE) model of bivalent antibody-antigen binding in the context of mAb cancer therapies, highlighting the added complexity associated with bivalency of the antibody. To understand what drives the complex temporal dynamics of bivalent antibody-antigen binding, we construct asymptotic approximations to the model's solutions at different timescales and antibody concentrations that are in good agreement with numerical simulations of the full model. We show how the dynamics differ between two scenarios; a region where unbound antigens are abundant, and one where the number of unbound antigens is small such that the dominant balance within the model equations changes. Of particular importance to the potency and efficacy of mAb treatments are the values of quantities such as antigen occupancy and bound antibody number. We use the results of our asymptotic analysis to approximate the long-time values of these quantities that could be combined with experimental data to facilitate parameter estimation.}




\maketitle

\section{Introduction}\label{sec1}

Ligand-receptor interactions are ubiquitous in biology and medicine \citep{Bongrand1999}. A ligand, often a small molecule or protein, binds specifically to a receptor protein on the surface of, or within, a cell \citep{Teif2005}. Many processes within the body are initiated by a ligand binding to its receptor. For example, enzymes bind to substrates to catalyse biochemical reactions while hormones bind to their receptors to regulate physiology \citep{Attie1995}.\par
A key ligand-receptor interaction is an antibody binding to its target antigen \citep{Goldberg1952}. Antibodies play an important role in adaptive immunity by binding to harmful substances such as toxins to neutralise them \citep{Forthal2014}. A type of antibody-based immunotherapy is based on monoclonal antibodies (mAbs). MAbs are used to treat many pathologies, including Alzheimer's, autoimmune diseases, and cancer \citep{Adams2005, Hafeez2018, Cummings2023}. \par
MAbs can be used to target cancer in a variety of ways. For example, immune checkpoint inhibitors bind to checkpoint receptors on a tumour cell surface, mitigating immune response inhibition \citep{Shiravand2022}. Core to the mechanism of action of immune checkpoint inhibitors is the antibody binding to its target antigens on the tumour cell's surface \citep{Junker2021}. In particular, binding increased numbers of immune checkpoint receptors and thus increasing antigen occupancy, reduces the ability of a tumour cell to engage these receptors with lymphocytes such as CD8 and CD4 T cells and in so doing, down-regulate the immune response. Additionally, levels of bound antibody have been shown to influence the efficacy and potency of antibody effector functions such as antibody-dependent cellular cytotoxicity (ADCC) \mbox{\citep{Mazor2016, Junker2021}}.\par
While many ligands are monovalent, the ability of a mAb to bind its target antigens is greatly enhanced by the fact that most mAbs have two antigen binding arms and thus are are bivalent \citep{Vauquelin2013}. The binding of multiple antigens decreases the effective dissociation between mAb and cell leading to an apparent increase in binding strength, termed the ``avidity effect" \citep{Oostindie2022}. \par 
Clearly, antibody-target antigen binding dynamics are important for the potency and efficacy of mAb therapies and are complicated by bivalency of the antibody. However, to the best of our knowledge, a detailed mathematical analysis of these binding dynamics in the context of cancer therapies and the avidity effect has not been completed. In this work, we will utilise asymptotic analysis to investigate how antibody-antigen binding dynamics change across timescales and antibody concentrations.\par
Mathematical modelling has been used to study ligand-receptor interactions in a variety of contexts \citep{Perelson1980, Klotz1984, Goldstein2004, Finlay2020, Dey2023}. Ordinary differential equation (ODE) models have been used to study antibody-antigen interactions due to their simplicity \citep{Kaufman1992, Rhoden2016, Sengers2016, Heirene2024}. Of these studies, \cite{Perelson1980} has provided a detailed analysis of an ODE model of bivalent ligand receptor binding when the ligand is in excess. Here, we extend their analysis to the context of an antibody binding to antigens where the antibody concentration ranges over multiple orders of magnitude. Furthermore, due to the diffusion of antigens on the cell surface, binding of the second arm of the antibody is predicted to occur on a short timescale \citep{Sengers2016, Heirene2024}. Consequently, multiple timescales are present, and thus asymptotic analysis is a natural tool to characterise the binding dynamics. \par
In previous work, we used mathematical modelling to determine how antibody-antigen interactions affect the avidity effect, and equilibrium values of antigen occupancy and bound antibody numbers; these quantities are known to affect the potency and efficacy of immune checkpoint inhibitors and mAb effector functions, respectively \citep{Heirene2024}. We used global parameter sensitivity analysis to establish relationships between model parameters and antigen occupancy and bound antibody numbers. We found that the parameter dependencies were dose-dependent, with model outputs only becoming sensitive to binding parameters, such as the off rate at high antibody concentrations. However, in our previous work, we assumed the antibody-antigen interactions had reached equilibrium. Here, we will instead consider antibody-antigen binding dynamics and use asymptotic analysis to address two main aims. The temporal dynamics of bivalent antibody-antigen binding is complex due to the reactivity of the second antibody binding arm. Our first aim is, therefore, to use asymptotic analysis to elucidate the mechanisms that drive antibody-antigen binding on different timescales and across a range of antibody concentrations that commonly arise within in vitro experiments. We focus on in vitro experiments because they are widely used during the pre-clinical development of mAbs and can be more easily compared to the output of a simple mathematical model. Secondly, we aim to use asymptotic analysis to identify those model parameters that determine the long-time behaviour of quantities that impact mAb potency and efficacy. In particular, we seek to approximate expressions for antigen occupancy and bound antibody numbers.\par 
The remainder of the paper is organised as follows. In Section \ref{sec2} we formulate and nondimensionalise our model of bivalent antibody-antigen binding. In Section \ref{sec_dynamics} we show how the qualitative dynamics change with antibody concentration which we then analyse using perturbation methods in Section \ref{sec: biv}. The paper concludes in Section \ref{sec: discussion} where we summarise and discuss our results.
\section{Mathematical Model}\label{sec2}
\begin{figure}[http]
    \centering
    \includegraphics[width=1\textwidth]{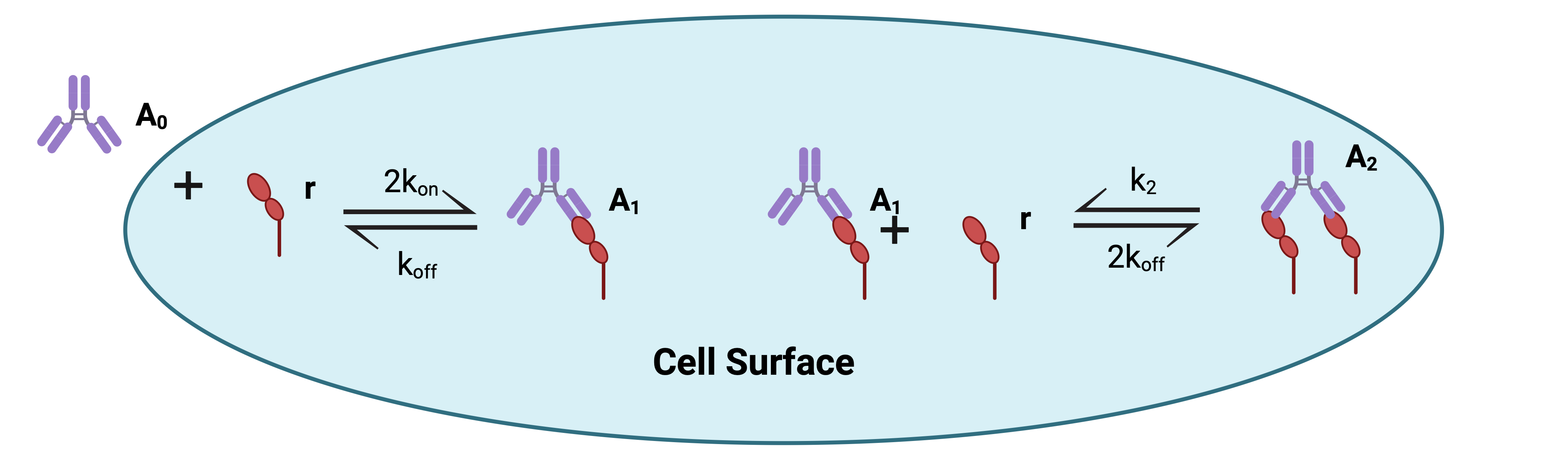}
    \caption{Schematic of a bivalent, monospecific antibody binding a target antigen on a cell surface. An unbound antibody, $A_0$, binds reversibly with a free target antigen, $r$, at a rate $k_{\text{on}}$, to form a monovalently bound antibody $A_1$ and dissociates at a rate $k_{\text{off}}$. $A_1$ binds a second free antigen at a rate $k_2$, to form a bivalently bound antibody $A_2$. A bivalently bound antibody can dissociate one of its bound arms away from the target antigen at a rate $k_{\text{off}}$. The factor of 2 appears in front of reaction rates where two antibody arms can undertake that reaction. In a slight abuse of notation, $A_0$ denotes a single unbound antibody but in Equations (\ref{model1})-(\ref{model4}) it denotes the total number of unbound antibodies (similarly for all other variables). Created with \href{https://www.biorender.com}{Biorender.com.}}
    \label{fig:Figure 2}
\end{figure}
In this section, we introduce a time-dependent mathematical model that describes the binding of a monospecific, bivalent antibody to target antigens on the surface of a single cell. Our model is based on an existing model of bivalent ligand-receptor binding presented in \cite{Perelson1980} and was first presented in \mbox{\cite{Heirene2024}}. \par
We apply the law of mass action to the reaction scheme shown in Figure \ref{fig:Figure 2} to derive a system of ODEs that describe the time evolution of the following dependent variables: the number of unbound target antigens, $r(t)$; the number of unbound antibodies, $A_0(t)$;  the number of monovalently bound antibodies, $A_1(t)$; and the number of bivalently bound antibodies, $A_2(t)$ \citep{Voit2015}. 
We assume that the system is well mixed and that target antigens are distributed uniformly over the cell surface. For simplicity, we neglect antigen internalisation. This is justified because, typically, the timescale of antigen internalisation is much slower than that of antibody binding on the surface of a cell \citep{Birtwistle2009, Vainshtein2014}.\par 
We assume a free monospecific antibody binds one of its arms to a free target antigen to form a monovalently bound antibody-antigen complex. The monovalently bound antibody can either dissociate its antigen-bound arm or bind a second free antigen with its unbound arm to form a bivalently bound antibody. Each antigen-bound arm may then dissociate from its antigen. The rate at which one arm dissociates is assumed to be equal to, and independent of, the rate at which the other arm dissociates. \par
Under the above assumptions and following the model development presented in \cite{Heirene2024}, our mathematical model can be written as follows
\begin{align}
    \frac{\mathrm{d}r}{\mathrm{d}t} &= -2k_1rA_0 + k_{\text{off}}A_1 - k_2rA_1 + 2k_{\text{off}}A_2, \label{model1}\\
    \frac{\mathrm{d}A_0}{\mathrm{d}t} &= -2k_1rA_0 + k_{\text{off}}A_1,\label{model2} \\
    \frac{\mathrm{d}A_1}{\mathrm{d}t} &= 2k_1rA_0 - k_{\text{off}}A_1 - k_2rA_1 + 2k_{\text{off}}A_2, \label{model3}\\
    \frac{\mathrm{d}A_2}{\mathrm{d}t} &= k_2rA_1 - 2k_{\text{off}}A_2\label{model4}, 
\end{align}
 where the reaction rates are defined and described in Figure {\ref{fig:Figure 2}}. The factor of 2 that appears in the reaction terms $2k_1rA_0$ and $2k_{\text{off}}A_2$ arises when two antibody arms can undertake a reaction (e.g. an antibody in solution can bind either of its arms and, similarly, when it is bivalently bound either arm may dissociate). Equations (\ref{model1})-(\ref{model4}) are closed by imposing the following initial conditions:
 \begin{equation}
     r(0)=r_{\text{tot}}, \hspace{2mm} A_0(0)=A_{\text{tot}}, \hspace{2mm}  A_1(0)=0, \hspace{2mm}  A_2(0)=0. \label{ics}
 \end{equation}
In Equation (\ref{ics}), we assume that all antigens are initially unbound, and we denote by $A_{\text{tot}}$ and $r_{\text{tot}}$ the total number of antibodies and target antigens respectively. The units of $A_{\text{tot}}$ and $r_{\text{tot}}$ are antibody number and antigen number respectively.\par
By taking suitable linear combinations of Equations (\ref{model1})-(\ref{model4}) and utilising Equation (\ref{ics}), it is straightforward to show that the total number of antibodies and antigens are conserved quantities within the system:
\begin{align}
    A_0 + A_1 + A_2 &= A_{\text{tot}},\label{A conserv} \\
    r + A_1 + 2A_2 &= r_{\text{tot}}\label{r conserv}.
\end{align}
We use Equations (\ref{A conserv}) and (\ref{r conserv}) to eliminate $A_0 =A_{\text{tot}} -  A_1 - A_2$ and $r = r_{\text{tot}} -A_1 - 2A_2$ and, henceforth, focus on the following reduced system for $A_1(t)$ and $A_2(t)$:
\begin{align}
    \begin{split}
        \frac{\mathrm{d}A_1}{\mathrm{d}t} &= 2k_1(r_{\text{tot}}-A_1-2A_2)(A_{\text{tot}}-A_1-A_2) - k_{\text{off}}A_1 \\
       & - k_2(r_{\text{tot}}-A_1-2A_2)A_1 +  2k_{\text{off}}A_2,
    \end{split}
        \label{reduc eqn A1} \\
    \frac{\mathrm{d}A_2}{\mathrm{d}t} &=  k_2(r_{\text{tot}}-A_1-2A_2)A_1 -  2k_{\text{off}}A_2, \label{reduc eqn A2}
\end{align}
with
\begin{equation}
    A_1(0)=A_2(0)=0.\label{reduc ics}
\end{equation}
Before nondimensionalising our reduced model, we pause to estimate the model parameters.
\subsection{Model Parameter Estimates}
Model parameter estimation was presented in \cite{Heirene2024} and is summarised here for completeness. Noting that assays within a reaction volume, $V_{\text{well}}$, (units: litres L), are used to estimate parameters, we estimate $A_{\text{tot}}$, the number of antibodies within the system, for a given experimental antibody concentration from
\begin{equation}
    A_{\text{tot}} = A_{\text{init}}\sigma. \label{Atot_eqn}
\end{equation}
Here, $A_{\text{init}}$ is the initial antibody concentration (units: molar concentration, M = mol/L), and
$\sigma$ is a ``concentration-to-antibody-number" conversion factor given by
\begin{equation}
    \sigma = \frac{V_{\text{well}}Na}{T^0}, \label{sigma eqn}
\end{equation}
where  $Na=6.02214 \times 10^{23}$ is Avogadro's number (units: $\text{mol}^{-1}$) and $T^0$ is the number of target cells within the reaction volume. Equation (\ref{sigma eqn}) is normalised with respect to $T^0$, because we are focusing on binding to a single target cell. \par
Estimates of $k_{\text{on}}$, the in-solution binding rate (units s$^{-1}$M$^{-1}$), are given for mAbs in the literature \citep{Bostrom2011, Mazor2016}. Here, however, we consider numbers of antibodies and antigens, rather than concentrations. Therefore, we need to rescale $k_{\text{on}}$ so its units are compatible with those used for $A_1$ and $A_2$:
\begin{equation}
    k_1 = \frac{k_{\text{on}}}{\sigma},\label{k1}
\end{equation}
where $k_1$ is in units of the number of antibodies per second.\par 
As detailed in \cite{Heirene2024}, we assume that binding of the second arm of the antibody to antigens on the cell surface is limited by antigen diffusion \mbox{\citep{Sengers2016}}. This is because in our previous work we showed that after an antibody binds a single antigen, it is unlikely that a second antigen will be close enough to bind the antibody's second arm. We assume that binding of the second arm is driven by the antigen's ability to diffuse across the cell surface and get close enough to bind the antibody. From a consideration of first passage time processes (\cite{Heirene2024} and references therein as well as \cite{Sengers2016}), we suppose that this reaction occurs at rate $k_2$ where
\begin{equation}
    k_2 = \frac{D}{4\pi(T_{\text{rad}})^2}. \label{Sk2}
\end{equation}
In Equation (\ref{Sk2}), $D$ is the diffusion coefficient of the target antigen (units: $\text{m}^2\text{s}^{-1}$) and $T_{\text{rad}}$ is the radius of the target cell (units: metres). For reference, the model parameters and their interpretations are provided in Table \ref{param table}.\par

\begin{sidewaystable}
\centering
\caption{Model parameters associated with Equations (\ref{model1})-(\ref{model4})}\label{param table}
\begin{tabular}{@{}llll@{}} 
\toprule
Parameter & Definition & Estimated Value (units) & Source \\
\midrule
$r_{\text{tot}}$ & Target cell antigen number & $10^4 - 10^6$ (receptors per cell) & \cite{Mazor2016}\\ 
\hline
$k_{\text{on}}$ &  Antibody in solution binding rate & $10^4 - 10^6$ (s$^{-1}$ M$^{-1}$)  &\cite{Bostrom2011, Mazor2016}\\
\hline
$k_{\text{off}}$ &  Antibody dissociation rate & $10^{-6}-10^{-3}$ (s$^{-1}$) & \cite{Bostrom2011, Mazor2016}\\
\hline
$A_{\text{init}}$ & Initial antibody concentration & $10^{-12} - 10^{-5}$ (M) & \cite{Pollard2010} \\
\hline
$T^{0}$ & Target cell number in assay & $2\times 10^5$ (cells) & \cite{Yu2017}\\
\hline
$V_{\text{well}}$ & Assay reaction well volume & $150$ $(\mu$L)  & \cite{Yu2017}\\
\hline
$T_{\text{rad}}$ & Tumour cell radius & $8$ $(\mu$m) & \cite{Hosokawa2013}\\
\hline
$D$ & Target antigen membrane diffusion coefficient & $10^{-15}-10^{-13}$  (m$^2$s$^{-1}$) & \cite{McCloskey1986}\\
\hline
$k_2$ & Diffusion limited second arm binding rate & $10^{-6} - 10^{-4}$ (s$^{-1}$) & \cite{Sengers2016}\\
\hline
$\sigma = V_{\text{well}} Na/T^0$ & Concentration to protein number conversion factor & $4.5 \times 10^{14}$ (M$^{-1}$cell$^{-1}$)& \\
\hline
$ Na$ & Avogadro constant & $6.02214 \times 10^{232}$ (mol$^{-1}$)& \\
\botrule
\end{tabular}
\end{sidewaystable}

\subsubsection{Nondimensionalisation}
We nondimensionalise our model, by introducing the following scalings
\begin{equation}
    A_0 = A_{\text{tot}}\hat{A_0}, A_1=r_{\text{tot}}\hat{A_1}, A_2 = r_{\text{tot}}\hat{A_2}, r=r_{\text{tot}}\hat{r}, t=\hat{\tau}/k_{\text{off}} \label{nondim scalings}.
\end{equation}
Dropping the hats for notational simplicity, we arrive at the following dimensionless equations
\begin{align}
    \frac{\mathrm{d}A_1}{\mathrm{d}\tau} &= 2\alpha_1 (1-A_1-2A_2)(\beta - A_1 - A_2) - A_1 - \alpha_2(1-A_1-2A_2)A_1 + 2A_2, \label{nondim1} \\
    \frac{\mathrm{d}A_2}{\mathrm{d}\tau} &= \alpha_2(1-A_1-2A_2)A_1 - 2A_2, \label{nondim2}
\end{align}
with
\begin{equation}
    A_1(0)= A_2(0) = 0. \label{nondim ics}
\end{equation}
In Equations (\ref{nondim1}) and (\ref{nondim2}), we have introduced the following dimensionless parameter groupings
\begin{equation}
    \alpha_1 =  k_1r_{\text{tot}}/k_{\text{off}},
    \alpha_2 = k_2r_{\text{tot}}/k_{\text{off}}, 
    \beta = A_{\text{tot}}/r_{\text{tot}}. \label{nondim express}
\end{equation}
For reference, the nondimensional parameters are provided in Table \ref{nondim params}. We note that the binding of the second arm of the antibody on the cell surface is a fast reaction compared to the binding of the first arm ($\alpha_2 \gg \alpha_1)$. This drives the avidity effect where the increase in the number of bivalently bound antibody, $A_2$, contributes to an increase in binding strength because bivalently bound antibodies are less likely to dissociate. Physically, $\alpha_2 \gg \alpha_1$ is due to the proximity of free antigens to the antibody and the ability of antigens to diffuse across the cell surface. For $r_{\text{tot}}=10^5$, $k_{\text{on}}=10^5$ M$^{-1}$s$^{-1}$, $k_{\text{off}}=10^{-4}$ s$^{-1}$ and $k_2=10^{-5}$ s$^{-1}$, we estimate that $\alpha_2 = \text{ord}(10^4)$ and choose $\epsilon = 10^{-2}$ to be a small parameter such that $\alpha_2 = \hat{\alpha}_2/\epsilon^2$ (where $\hat{\alpha}_2$ is an $\text{ord}(1)$ quantity). Note that for all parameter combinations in Table {\ref{param table}} we can choose an $\epsilon \ll 1$ such that the value of $\alpha_2 = \text{ord}(\epsilon^{-2})$, where we introduce the notation $X=\text{ord}(Y)$ which means $X/Y$ is strictly of order unity as $\epsilon \rightarrow 0$ \mbox{\citep{Hinch1991}}. \par
Upon nondimensionalisation, Equations (\ref{A conserv}) and (\ref{r conserv}) become
\begin{align}
    1 &= A_0 +\frac{1}{\beta}\bigg(A_1 + A_2 \biggr), \label{nondim A conserv} \\
    1 &= r + A_1 + 2A_2. \label{nondim r conserv}
\end{align}
which give the constraints
\begin{align}
    A_1+A_2 \leq \beta, \label{nondim_conserv1} \\
    A_1+2A_2 \leq 1 \label{nondim_conserv2}.
\end{align}
These constraints state that the number of antibodies bound to the cell surface and antigens bound with antibody are bounded by the total number of antibodies within the system and the target antigen density respectively. We deduce further that antibodies will bind to the cell surface until either antibodies or free antigens run out.\par
In what follows, it will be convenient to recast Equations (\ref{nondim1}) and (\ref{nondim2}) in terms of $W = A_1 + A_2$ and $A= A_2$ in order to simplify algebra calculation. It is straightforward to show that $W$ and $A$ satisfy the following ODEs
\begin{align}
    \frac{\mathrm{d}W}{\mathrm{d}\tau} &= 2\alpha_1(1-W-A)(\beta-W) - (W-A), \label{W eqn}\\
    \frac{\mathrm{d}A}{\mathrm{d}\tau} &= \frac{2}{\delta}\biggl[(W-A)(1-W-A) -\delta A\biggr], \label{A eqn}
\end{align}
with
\begin{equation}
    W(0)=A(0)=0,
\end{equation}
 and where, for simplicity, we have defined
\begin{equation}
    \delta = \frac{2\epsilon^2}{\hat{\alpha_2}}. \label{delta}
\end{equation}
Note that since $\alpha_2 \gg \alpha_1$ we also have that
\begin{equation}
    \delta = \frac{2}{(\frac{\hat{\alpha}_2}{\epsilon^2})} \ll \frac{1}{\alpha_1}.
\end{equation}

\begin{table}[ht]
\centering
\caption{Model Nondimensional Parameters, $\alpha_1$ and $\alpha_2$ are constrained such that $\alpha_2 \gg \alpha_1$, so that the secondary binding is rapid compared to the primary binding.} \label{nondim params}
\begin{tabular*}{\textwidth}{@{}lll@{}} 
 \toprule
Parameter & Definition & Estimated Value (units) \\
 \midrule
$\alpha_1 = \frac{k_1r_{\text{tot}}}{k_{\text{off}}}$ & Nondimensional antibody in solution target binding rate & $2.2 \times 10^{-4}- 2.2 \times 10^{3}$ \\
  \hline
 $\alpha_2 = \frac{k_2r_{\text{tot}}}{k_{\text{off}}}$ & Nondimensional diffusion limited second arm binding rate & $50 - 5 \times 10^8 $ \\
  \hline
 $\beta = \frac{A_{\text{tot}}}{r_\text{tot}}$ & Ratio of Antibody to target receptor within the system & $5 \times 10^{-4} - 5 \times 10^5$ \\
 \botrule
\end{tabular*}
\centering
\end{table} 
\section{Bivalent Model Dynamics}\label{sec_dynamics}
In the previous section, we introduced the parameter $\beta=A_{\text{tot}}/r_{\text{tot}}$, the ratio of total antibody to antigen number. With $r_{\text{tot}} = 10^5$, a typical antigen density for a tumour cell \citep{Mazor2015}, $\beta$ ranges over the following orders of magnitude as the antibody concentration $A_{\text{init}}$ varies over a range commonly used within in vitro experiments:
\begin{enumerate}
    \item $A_{\text{init}}= 10^{-11} \text{ M} \Leftrightarrow \beta = \text{ord}(\epsilon)$
    \item $A_{\text{init}}= 10^{-9} \text{ M} \Leftrightarrow \beta = \text{ord}(1)$
    \item $A_{\text{init}}= 10^{-7} \text{ M}\Leftrightarrow \beta = \text{ord}(\epsilon^{-1})$
    \item $A_{\text{init}}= 10^{-5} \text{ M}\Leftrightarrow \beta = \text{ord}(\epsilon^{-2})$
\end{enumerate}
Note that if the value of $r_{\text{tot}}$ were to change, the same magnitudes of $\beta$ can be achieved by adjusting $A_{\text{init}}$. The only condition we require for the analysis that follows is $\alpha_2 = k_2r_{\text{tot}}/k_{\text{off}} \gg \alpha_1$ to account for the avidity effect. Since, $k_{\text{off}}/k_2 \leq 100$ and $r_{\text{tot}} \geq 10^4$ for parameter ranges reported in the literature, this condition is satisfied. \par
\begin{figure}[h!]\captionsetup[subfigure]{font=normal}
\captionsetup{width=1\textwidth}
\begin{tabular}{cccc}
\subfloat[$\beta = \text{ord}(\epsilon)$]{\includegraphics[width=0.5\linewidth]{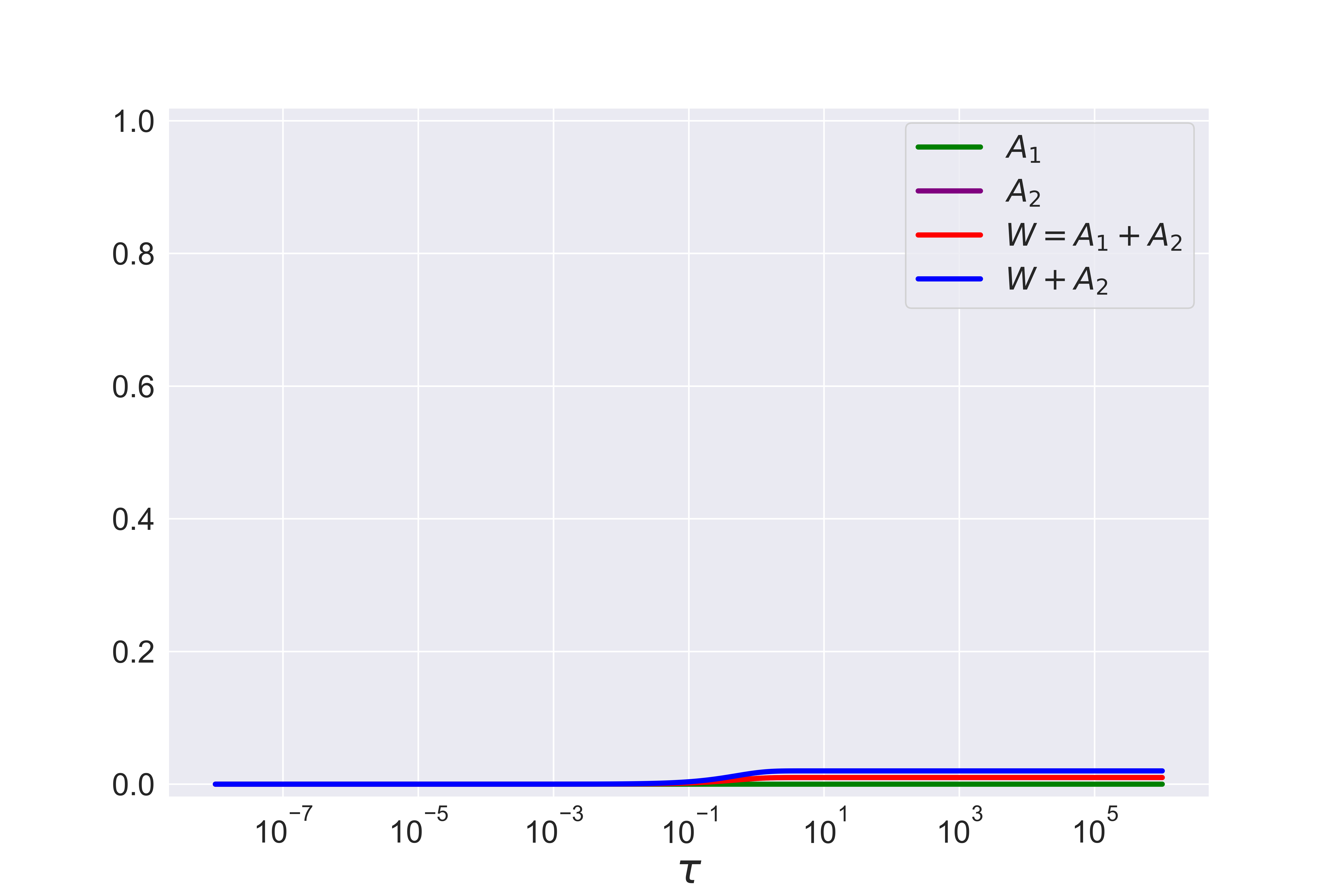}\label{sim_1e-2}} &
\subfloat[$\beta = \text{ord}(1)$] {\includegraphics[width=0.5\linewidth]{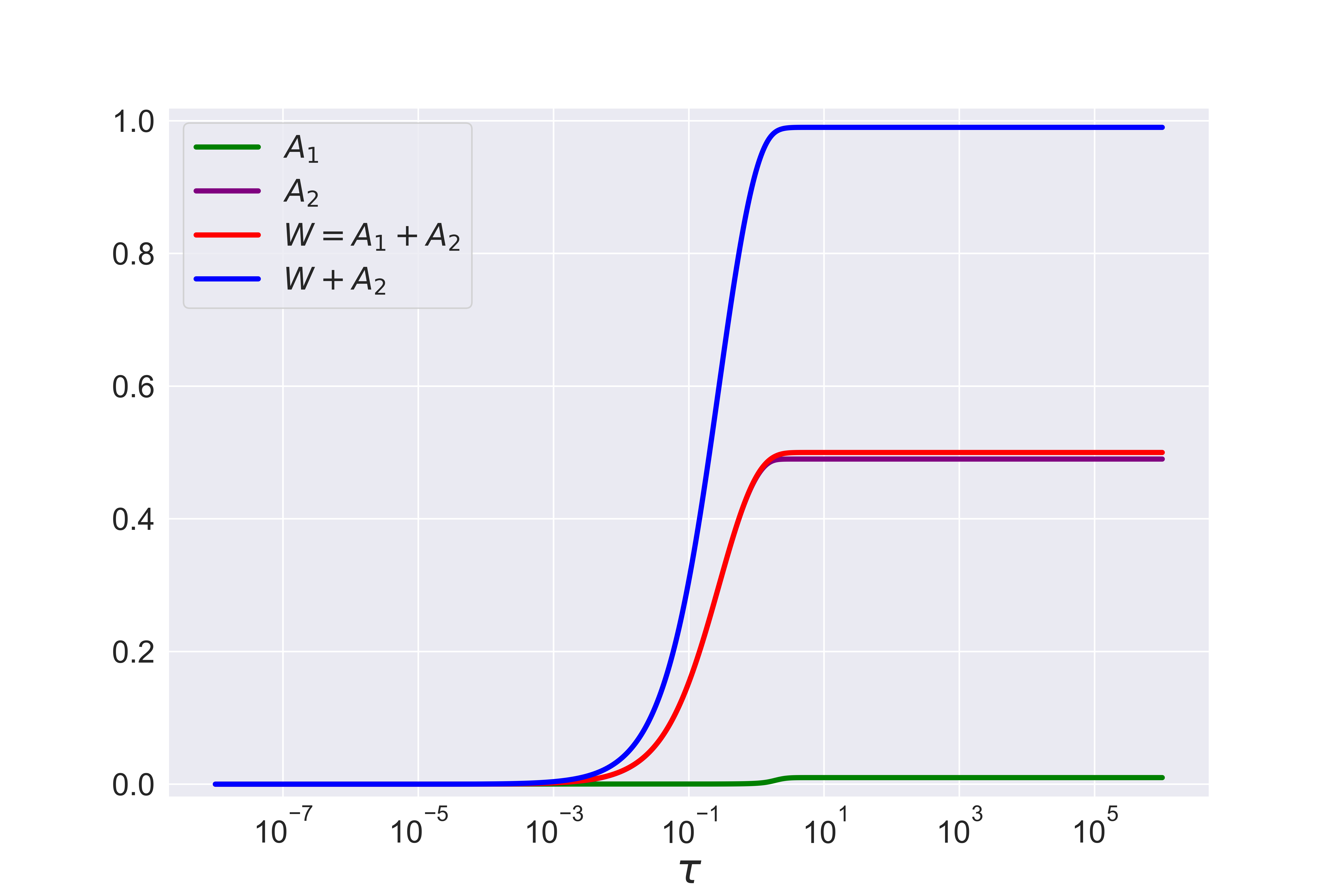}\label{sim_1}} \\
\subfloat[$\beta = \text{ord}(\epsilon^{-1})$]{\includegraphics[width=0.5\linewidth]{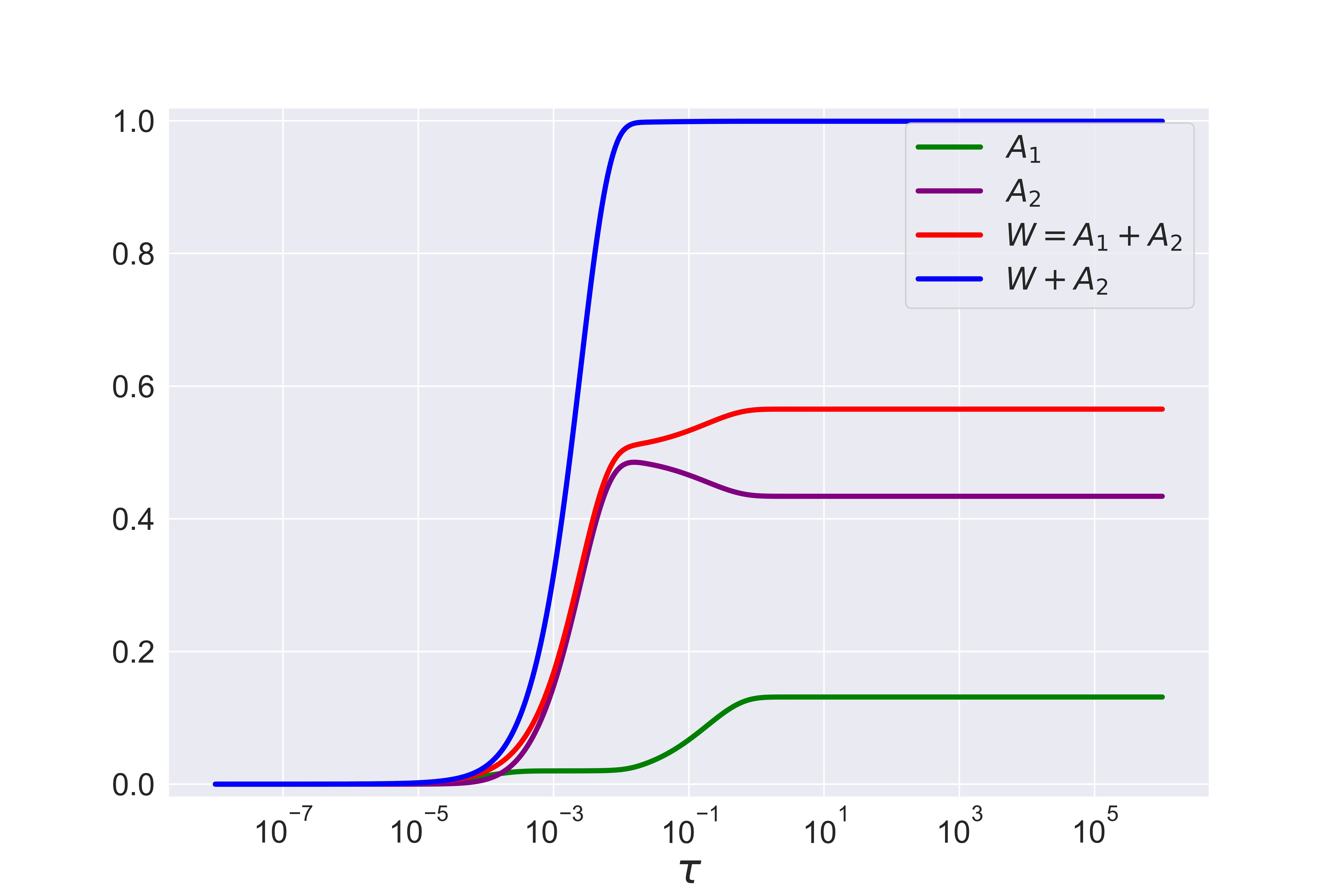}\label{sim_1e2}} &
\subfloat[$\beta = \text{ord}(\epsilon^{-2})$]{\includegraphics[width=0.5\linewidth]{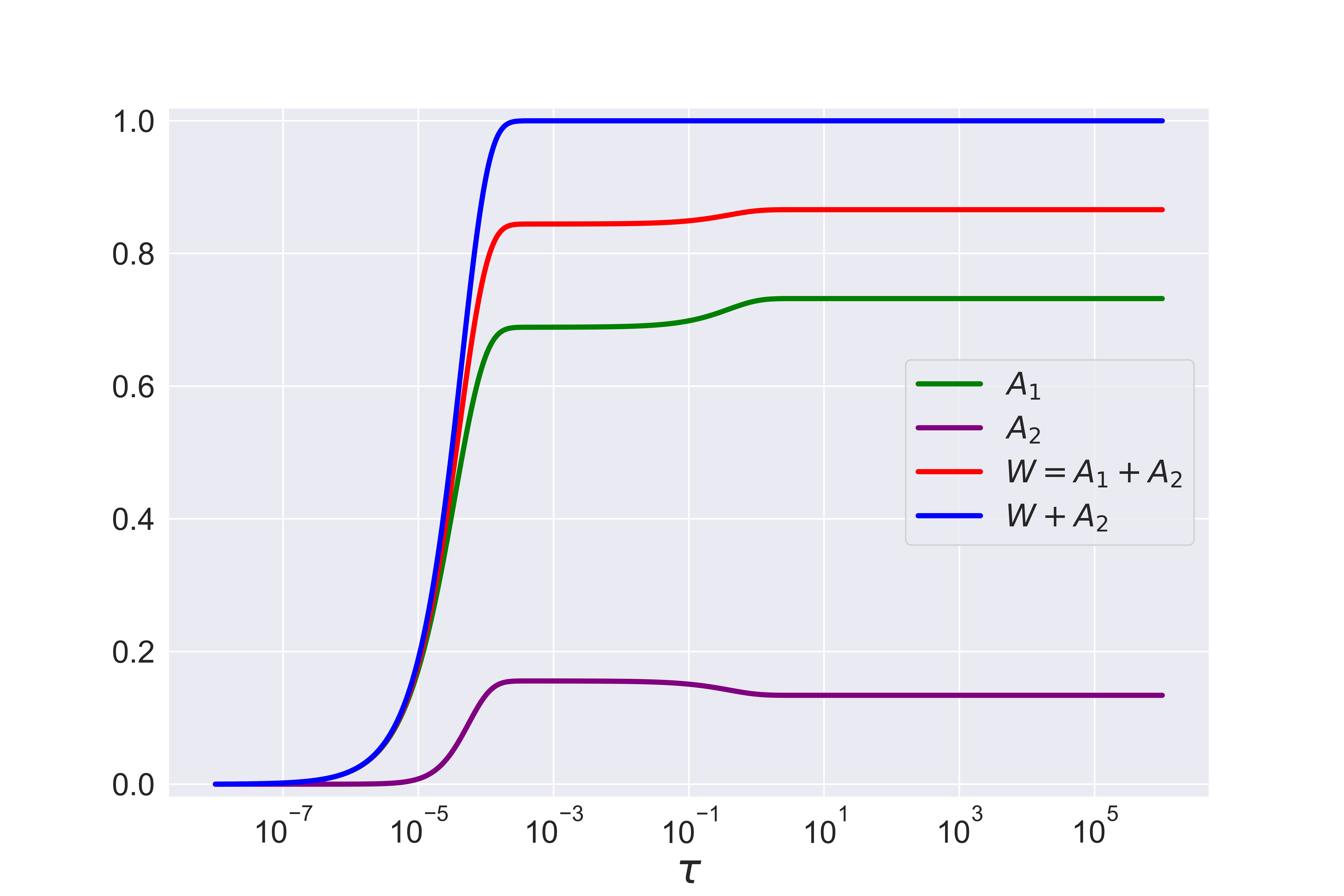}\label{sim_1e4}} \\
\end{tabular}
\centering
\caption{Model Dynamics from Equations (\ref{nondim1}) and (\ref{nondim2}) for different orders of magnitude of $\beta$ with $\alpha_1=1$, $\alpha_2 = 10^4$. Initial condition of the system is $A_1(0)=A_2(0)=0$.}
\label{fig: dynamics}
\end{figure}

 In Figure \ref{fig: dynamics}, we present simulations of Equations (\ref{nondim1}) and (\ref{nondim2}) as $\beta$ varies. We summarise the results below:
\begin{itemize}
    \item Figure \ref{sim_1e-2} ($\beta = \text{ord}(\epsilon))$. Here, antigens are in excess since the number of antibodies is small. As a result, the total amount of binding is small, and $A_2$ reaches a maximum value of $A_2=0.01$. There are fewer monovalently bound antibodies present because free antigens are in excess; once an antibody binds one arm, its second arm also binds as it is not crowded out.
    \item Figure \ref{sim_1} ($\beta = \text{ord}(1)$). Here, antibody and antigen levels are similar, so all antigens are saturated with antibody. Again, there are very few monovalently bound antibodies, but their numbers increase slightly at longer timescales as bivalently bound antibodies dissociate one of their arms. Similarly, competition for binding sites remains small because the magnitude of the reaction rate associated with the second binding event is large ($\alpha_2 \gg \alpha_1$) so, as for Figure \ref{sim_1e-2}, $A_2$ dominates $A_1$.
    \item Figure \ref{sim_1e2} ($\beta = \text{ord}(\epsilon^{-1})$). In this regime, antibody is in excess. However, despite increased competition for binding sites, since the reaction rate of the second binding event is much larger than that of the first binding event ($\alpha_2 \gg \alpha_1$), the surface reaction ``out-competes" the in-solution reaction and $A_2$ dominates at short timescales ($\tau =\text{ord}(\epsilon)$). When $\tau = \text{ord}(1)$, however, one arm of the bivalently bound complex dissociates and the free antigen is quickly bound by a free antibody. Consequently, the number of bivalently bound antibodies decreases and the number of monovalently bound antibodies increases. 
    \item Figure \ref{sim_1e4} ($\beta = \text{ord}(\epsilon^{-2})$). Here, the large number of antibodies compete for a smaller number of free antigens, driving an effective increase in the rate at which the first binding event takes place so that it out-competes the fast surface reaction that forms $A_2$ and $A_1$ dominates. A separation of timescales is visible when $\beta = \text{ord}(\epsilon^{-2})$: the initial increase in $A_1$ ($\tau = \text{ord}(\epsilon^3)$) is followed by a smaller rise in $A_2$ ($\tau = \text{ord}(\epsilon^2)$), as the second arm binds, before dissociation events drive another increase in $A_1$ ($\tau = \text{ord}(1)$).
\end{itemize}

\section{Asymptotic Analysis of Bivalent Model}\label{sec: biv}
In this section, we construct approximate solutions to Equations (\ref{W eqn}) and (\ref{A eqn}) and consider separately the cases $\beta = \text{ord}(\epsilon)$, $\beta = \text{ord}(1)$, $\beta = \text{ord}(\epsilon^{-1})$ and $\beta = \text{ord}(\epsilon^{-2})$. 
\subsection{Perturbation Analysis for $\beta = \text{ord}(\epsilon)$}
\begin{figure}[h!]
    \centering
    \includegraphics[width=1\textwidth]{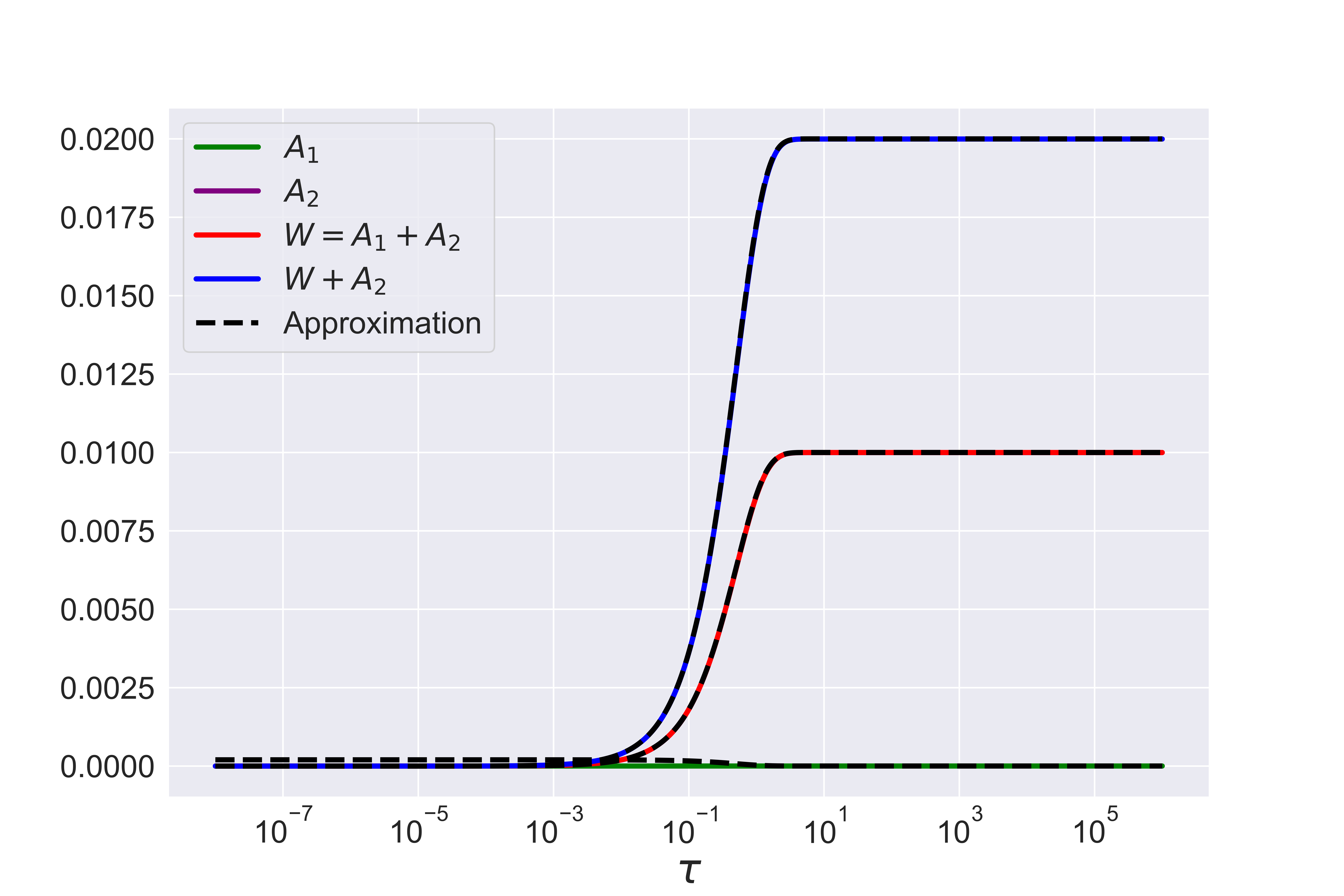}
    \caption{Result of asymptotic analysis of Equations (\ref{W epsilon}) and (\ref{A epsilon}) for $\beta = \text{ord}(\epsilon)$. The asymptotic approximations to the model dynamics are given by Equations (\ref{W epsilon approx})-(\ref{A1 epsilon approx}) and are denoted with dashed black lines. Numerical solutions of Equations (\ref{W epsilon}) and (\ref{A epsilon}) are denoted by the coloured solid lines.}
    \label{fig:1e-2 summary}
\end{figure}

With $\beta = \text{ord}(\epsilon)$, Equations (\ref{W eqn}) and (\ref{A eqn}) supply
\begin{align}
    \frac{\mathrm{d}W}{\mathrm{d}\tau} &= 2\alpha_1(1-W-A)(\epsilon \hat{\beta}-W) - (W-A), \label{W epsilon}\\
    \frac{\mathrm{d}A}{\mathrm{d}\tau} &= \frac{2}{\delta}\biggl[(W-A)(1-W-A) -\delta A\biggr], \label{A epsilon}
\end{align}
where  $\beta=\epsilon \hat{\beta}$, $\hat{\beta} = \text{ord}(1)$.
We rescale $W= \epsilon \bar{W}$ and $A= \epsilon \bar{A}$ because antigens are in excess of antibody. The governing equations then become
\begin{align}
    \frac{\mathrm{d}\bar{W}}{\mathrm{d}\tau} &= 2\alpha_1(1-\epsilon(\bar{W}+\bar{A}))( \hat{\beta}-\bar{W}) - (\bar{W}-\bar{A}), \label{small W eqn}\\
    \frac{\mathrm{d}\bar{A}}{\mathrm{d}\tau} &= \frac{2}{\delta}\biggl[(\bar{W}-\bar{A})(1-\epsilon(\bar{W}-\bar{A})) -\delta\bar{A}\biggr]. \label{small A eqn}
\end{align}
We seek regular power series expansions of the form
\begin{align}
    \bar{W} &= \bar{W}_{0} + \epsilon \bar{W}_{1} + \epsilon^2\bar{W}_{2} + \cdots, \label{small W exp}\\ 
    \bar{A} & = \bar{A}_{0} + \epsilon \bar{A}_{1} + \epsilon^2\bar{A}_{2} + \cdots, \label{small A exp}
\end{align}
Noting that $\delta =\text{ord}(\epsilon^2)$, substituting Equations (\ref{small W exp}) and (\ref{small A exp}) for $W$ and $A$ in Equations (\ref{small W eqn}) and (\ref{small A eqn}) and equating to zero terms of $\mathcal{O}(1)$, we have that
\begin{align}
    \frac{\mathrm{d}\bar{W}_{0}}{\mathrm{d}\tau} &= 2\alpha_1(\hat{\beta} - \bar{W}_{0}) - (\bar{W}_{0}-\bar{A}_{0}), \label{W small leading}\\
    0 &= (\bar{W}_{0}-\bar{A}_{0}). \label{A small leading}
\end{align}
Equation (\ref{A small leading}) gives $\bar{W}_{0} = \bar{A}_{0}$, in which case Equation (\ref{W small leading}) supplies
\begin{equation}
     \frac{\mathrm{d}W_{0}}{\mathrm{d}\tau} = 2\alpha_1(\hat{\beta} - \bar{W}_{0}),
\end{equation}
with
\begin{equation}
   \bar{A}_{0} = \bar{W}_{0} = \hat{\beta}(1-e^{-2\alpha_1\tau}) \rightarrow \hat{\beta} \text{   as   } \tau \rightarrow \infty.
\end{equation}
In this regime, antibody is limiting; all antibodies  bind to available antigen at rate $2\alpha_1$ until there is no free antibody left. To determine $A_1 = W-A = \epsilon(\bar{W}-\bar{A})$ we consider higher order terms when substituting Equations (\ref{small W exp}) and (\ref{small A exp}) into Equations (\ref{small W eqn}) and (\ref{small A eqn}). Comparing $\mathcal{O}(\epsilon)$ terms, we obtain $\bar{W}_{1}=\bar{A}_{1}$ with
\begin{equation}
    \bar{W}_{1}=\bar{A}_{1} = 2\hat{\beta}e^{-2\alpha_1\tau}\biggl(1 - 2\alpha_1\biggl[\tau + \frac{e^{-2\alpha_1\tau}}{2\alpha_1}\biggr]\biggr).
\end{equation}
Considering $\mathcal{O}(\epsilon^2)$ terms supplies
\begin{equation}
    \bar{A}_{2} = \bar{W}_{2} - \frac{2}{\hat{\alpha_2}}\biggl(\hat{\beta}(1-e^{-2\alpha_1\tau}) + \alpha_1e^{-2\alpha_1\tau}\biggr), \label{small 2nd order exp}
\end{equation}
Noting that the non-zero leading order expression for $A_1$ is thus given by $\epsilon^3(\bar{W}_{2} - \bar{A}_{2})$, Equation (\ref{small 2nd order exp}) supplies
\begin{equation}
    \bar{W}_{2} - \bar{A}_{2} = \frac{2}{\hat{\alpha_2}}\biggl(\hat{\beta}(1-e^{-2\alpha_1\tau})+\alpha_1 e^{-2\alpha_1\tau}\biggl).
\end{equation}
Hence, $A_1\rightarrow 2\epsilon^3\hat{\beta}/\hat{\alpha_2}$ as $\tau \rightarrow \infty$. To summarise, the solutions when $\beta = \text{ord}(\epsilon)$ are
\begin{align}
    W =A  &=  \beta(1-e^{-2\alpha_1\tau}) + 2\epsilon \beta e^{-2\alpha_1\tau}\biggl(1 - 2\alpha_1\biggl[\tau + \frac{e^{-2\alpha_1\tau}}{2\alpha_1}\biggr]\biggr) +\mathcal{O}(\epsilon^3), \label{W epsilon approx}\\
    A_1 &=  \frac{2\epsilon^2}{\hat{\alpha_2}}\biggl(\beta(1-e^{-2\alpha_1\tau})+\alpha_1e^{-2\alpha_1\tau}\biggl) +\mathcal{O}(\epsilon^3).\label{A1 epsilon approx}
\end{align}
We conclude that when $\beta = \text{ord}(\epsilon)$ there is low mAb treatment potency and efficacy due to the small numbers of bound antibody, $W$, and monovalently bound antibody, $A_1$. This is not surprising due to the small antibody concentration within the system. Furthermore, there will be a strong avidity effect as the bound antibodies are primarily bivalently bound. Figure {{\ref{fig:1e-2 summary}}} shows that there is  good agreement between the numerical solutions of Equations ({\ref{W eqn}}) and ({\ref{A eqn}}) and the results of our asymptotic analysis, given by Equations ({{\ref{W epsilon approx}}})-({{\ref{A1 epsilon approx}}}).

\subsection{Perturbation Analysis for $\beta = \text{ord}(1)$}
There are three cases to consider when $\beta = \text{ord}(1)$:
\begin{itemize}
    \item $\beta < 1/2$: The number of antibodies is slightly less than the number required to bind all antigens;
    \item  $\beta=1/2$: There are exactly enough antibodies to bind all antigens;
    \item $\beta > 1/2$: There are slightly more antibodies than the number required to bind all antigens.
\end{itemize}
This trichotomy arises because the antibodies are bivalent (i.e each one can bind two antigens) and, therefore, $\beta=1/2$ is the smallest value at which all antigens are able to be bound. In this section, we will first consider inner and outer solutions of the dynamics in order to show that after initial transients, $A$ can be taken to be at quasi-steady state.\par
For $\beta=\text{ord}(1)$, the perturbation problem is given by 
\begin{align}
    \frac{\mathrm{d}W}{\mathrm{d}\tau} &= 2\alpha_1(1-W-A)(\beta-W) - (W-A), \label{W eqn1}\\
    \frac{\mathrm{d}A}{\mathrm{d}\tau} &= \frac{2}{\delta}\biggl[(W-A)(1-W-A) -\delta A \biggr], \label{A eqn1}
\end{align}
where $\delta=\text{ord}(\epsilon^2)$. Noting that $W(0)=A(0)=0$, at sufficiently small $\tau$, $W$ and $A$ will be small such that $W \ll 1, A\ll 1$ and Equation ({\ref{W eqn1}}) for sufficiently early time supplies
\begin{equation}
   \frac{\mathrm{d}W}{\mathrm{d}\tau} \approx 2 \alpha_1 \beta,
\end{equation}
which has the solution
\begin{equation}
    W \approx 2\alpha_1\beta \tau + \mathcal{O}(\tau^2) \label{W inner sol}.
\end{equation}
Now, we can solve for $A(\tau)$ in Equation ({\ref{A eqn1}}) which, when $W \ll 1, A\ll 1$, reduces to
\begin{equation}
    \frac{\mathrm{d}A}{\mathrm{d}\tau} = \frac{2}{\delta}\biggl[W - A \biggr],
\end{equation}
which has the solution
\begin{equation*}
    A(\tau) = \frac{2}{\delta}e^{-\frac{2}{\delta}\tau}\int_0^{\tau}W(p)e^{\frac{2}{\delta}p}\mathrm{d}p,
\end{equation*}
\begin{equation}
\Rightarrow A(\tau) =W(\tau) \biggl[1 - e^{-\frac{2}{\delta}\tau}\biggl](1 + \mathcal{O}(\delta)). \label{A qss eqn}
\end{equation}
In the second line we have used Laplace's method to approximate the integral. From {(\ref{A qss eqn})}, and provided $\tau\gg \delta/2$, we have that $A$ reaches the following quasi-steady state on this timescale:
\begin{equation}
    A \approx W(\tau),
\end{equation}
where we also have $\tau \ll 1$ so that the approximations $W, A \ll1$ are valid. As $\tau$ increases such that $A$ and $W$ are no longer much less than one, we now analyse the following equations (having identified above that $A$ reaches a quasi-steady state on a fast timescale, we set $\mathrm{d}A/\mathrm{d}\tau=0$):
\begin{align}
    \frac{\mathrm{d}W}{\mathrm{d}\tau} &= 2\alpha_1(1-W-A)(\beta-W) - (W-A),\label{Region II 1} \\
    0 &= (W-A)(1-W-A) -\delta A \label{Region II 2}.
\end{align}
To proceed, we consider the roots of
\begin{equation}
    P(A) \coloneqq (W-A)(1-W-A) -\delta A,
\end{equation} 
These are given by the two branches
\begin{equation}
    A^{\pm} = \frac{1}{2}\biggl[(1+\delta) \pm \sqrt{(1-2W)^2 +2\delta + \delta^2} \biggr].  \label{sol branches}
\end{equation}
Noting $A\leq1$, only $A^-$ is compatible with the dynamics.\par
In general we have that $A\rightarrow A^-(W)$, which may inferred from $A\approx W$ at early time. To see this more formally, note that for $\tau>0$ the value of $W$ (which excludes $W=0$ at $\tau=0$) is such that  $W(1-W) \in (0,1)$ and $A \in [0,1]$. Considering the structure of $P(A)$ for this value of $W$, we have $P(0)>0$, $P^\prime(A^+)>0$, $P^\prime(A^-)<0$ with $P(A)$ quadratic, noting $W$ is considered fixed in $P(A)$ as it evolves only on the slow timescales. Since $A(\tau=0) \in [0,1]$ and $W>0$ for $\tau>0$, this is sufficient to ensure that $A(\tau) \rightarrow A^-(\tau)$, so that apart from fast transients we have $A=A^-$.\par
\begin{figure}[h!]
    \centering
    \includegraphics[width=0.8\textwidth]{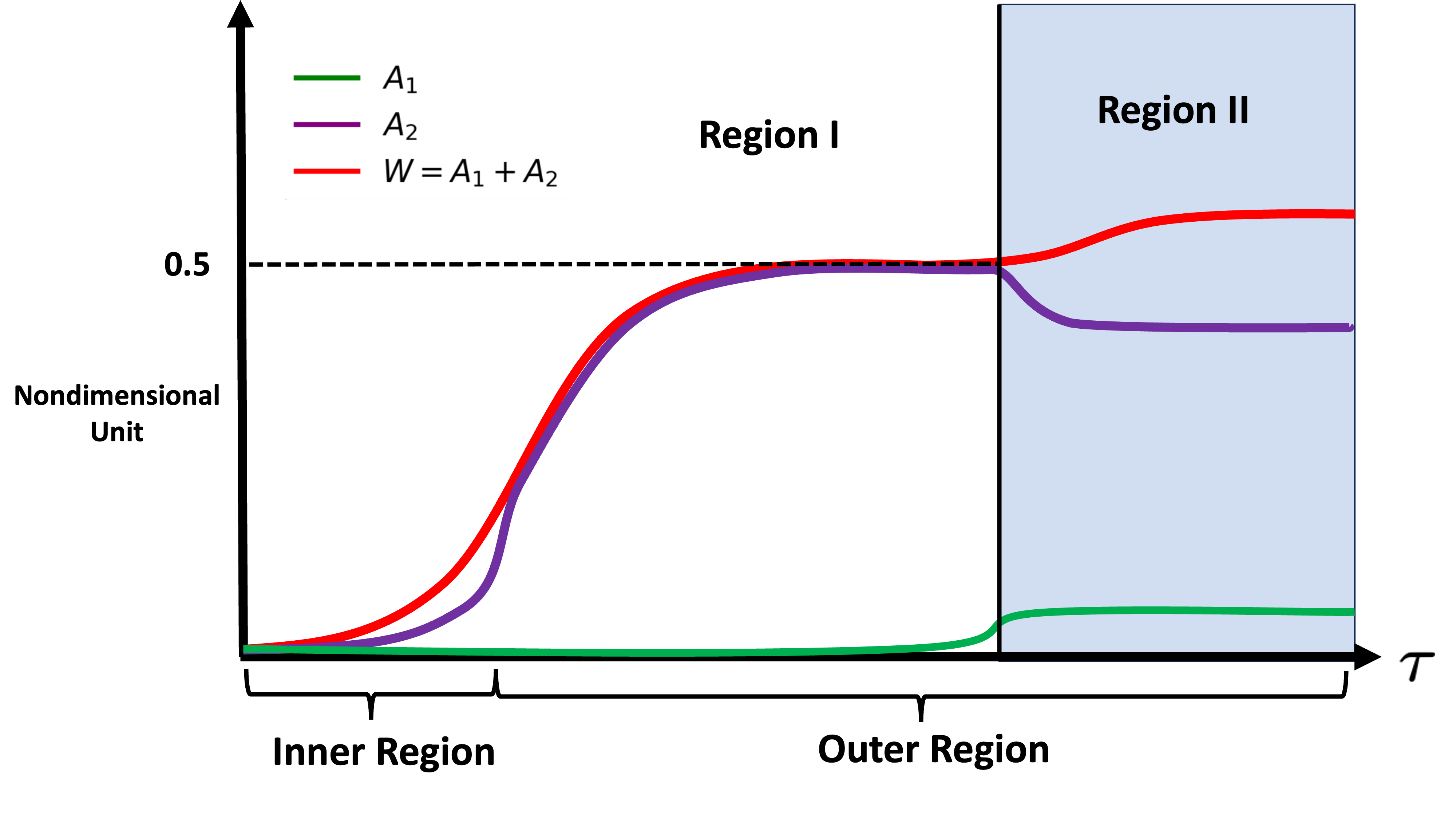}
    \caption{Figure depicting the inner and outer regions and Regions I and II within the dynamics. After the fast transients within the inner region, $A$ reaches a quasi equilibrium in the outer region defined by Equation (\ref{sol branches}). Region I is where there are many free antigens available while Region II is where the number of free antigens is small such that the size of terms in Equations (\ref{W eqn}) and (\ref{A eqn}) change and a new set of equations need to be derived to develop a simple approximation of the dynamics.}
    \label{fig: regionI/II }
\end{figure}

To find an expression $W$, we seek solutions to Equations (\ref{W eqn1}) and (\ref{A eqn1}) in the form of power series expansions in $\epsilon$, so that
\begin{align}
    W &= W_{\epsilon 0} + \epsilon W_{\epsilon1 } + \epsilon^2W_{\epsilon 2} + \cdots, \label{W exp} \\
    A &= A_{\epsilon 0} + \epsilon A_{\epsilon 1} + \epsilon^2A_{\epsilon 2} + \cdots. \label{A exp}
\end{align}
Substituting Equations (\ref{W exp}) and (\ref{A exp}) into Equations (\ref{W eqn1}) and (\ref{A eqn1}) and equating $\mathcal{O}(1)$ terms supplies
\begin{align}
    \frac{\mathrm{d}W_{\epsilon 0}}{\mathrm{d}\tau} &= 2\alpha_1(1 - W_{\epsilon 0} - A_{\epsilon 0})(\beta - W_{\epsilon 0}) - (W_{\epsilon 0} - A_{\epsilon 0}), \label{beta 1 W out} \\
    0 &= (W_{\epsilon 0} -A_{\epsilon 0})(1-W_{\epsilon 0}-A_{\epsilon 0}). \label{beta 1 A out}
\end{align}
Equation (\ref{beta 1 A out}) gives that either $W_{\epsilon 0} = A_{\epsilon 0}$ or $W_{\epsilon 0} + A_{\epsilon 0} =1$, though Equation (\ref{A qss eqn}) and the reasoning thereafter requires $W_{\epsilon 0}= A_{\epsilon 0}$. Then, Equation (\ref{beta 1 W out}) gives
\begin{equation}
    \frac{\mathrm{d}W_{\epsilon 0}}{\mathrm{d}\tau} = 2\alpha_1(1-2W_{\epsilon 0})(\beta - W_{\epsilon 0}) ,\label{beta 1 out first order}
\end{equation}
with solution when $\beta \neq 1/2$:
\begin{equation}
    W_{\epsilon 0}(\tau) = \frac{\beta(e^{2(2\beta - 1)\alpha_1 \tau} - 1)}{2\beta e^{2(2\beta - 1)\alpha_1 \tau} - 1}\rightarrow
    \begin{cases}
    \frac{1}{2} \text{ as } \tau \rightarrow \infty \text{ if } \beta >  \frac{1}{2}. \\
    \beta \text{ as } \tau \rightarrow \infty \text{ if } \beta < \frac{1}{2}.
    \end{cases}\label{beta neq 1/2 sol}
\end{equation}
When $\beta > 1/2$, there are enough antibodies to bind all antigens. As a result, antigens become saturated with bivalently bound antibody as $\tau \rightarrow \infty$. When $\beta < 1/2$, there are more antigens than antibody binding arms. Hence, all antibodies bind until there are no free antibodies as $\tau \rightarrow \infty$. 
When $\beta=1/2$, Equation (\ref{beta 1 out first order}) supplies
\begin{equation}
    W_{\epsilon 0}(\tau) = \frac{\alpha_1 \tau}{2\alpha_1\tau +1} \rightarrow \frac{1}{2} \text{ as } \tau \rightarrow \infty. \label{beta=1/2 sol}
\end{equation}
When $\beta \geq 1/2$ in Equations ({\ref{beta neq 1/2 sol}}) and ({\ref{beta=1/2 sol}}), our asymptotic expansion for $A_1 = W-A$ breaks down as $W \rightarrow 1/2$. To see this, upon substituting Equations ({\ref{W exp}}) and ({\ref{A exp}}) into Equation ({\ref{A eqn1}}) and equating terms of $\mathcal{O}(\epsilon)$ we have
\begin{equation}
    0 = (A_{\epsilon 0}-W_{\epsilon 0})(W_{\epsilon 1} + A_{\epsilon 1}) + W_{\epsilon 1} - A_{\epsilon 1},
\end{equation}
which, upon noting that $A_{\epsilon 0}=W_{\epsilon 0}$, gives $W_{\epsilon 1}=A_{\epsilon 1}$. Similarly, equating terms of $\mathcal{O}(\epsilon^2)$ gives
\begin{equation}
    \frac{\mathrm{d}W_{\epsilon 0}}{\mathrm{d}\tau} = \hat{\alpha}_2(W_{\epsilon 2}-A_{\epsilon 2})(1-2W_{\epsilon 0}) - 2W_{\epsilon 0},\label{higher order A}
\end{equation}
where we have substituted $W_{\epsilon 0}=A_{\epsilon 0}$. Upon rearranging Equation ({\ref{higher order A}}) for $W_{\epsilon 2}-A_{\epsilon 2}$, the leading order estimate of $A_1 = \epsilon^2(W_{\epsilon 2}-A_{\epsilon 2})$, we arrive at
\begin{equation}
    A_{1} \approx \frac{\epsilon^2}{\hat{\alpha}_2}\biggl[ \frac{\frac{\mathrm{d}W_{\epsilon 0}}{d\tau} + 2W_{\epsilon 0}}{1-2W_{\epsilon 0}}\biggr]. \label{blow up example}
\end{equation}
From Equation ({\ref{blow up example}}) we see that as $W_{\epsilon 0}\rightarrow1/2$ such that $|1-2W_{\epsilon 0}|\approx \mathcal{O}(\epsilon^2)$ in Equations ({\ref{beta neq 1/2 sol}}) and ({\ref{beta=1/2 sol}}), our estimate of $A_1$ breaks down. The above analysis highlights the existence of two regions within the dynamics when $\beta>1/2$:
\begin{itemize}
    \item Region I: Here the number of unbound antigens is large enough that the dominant balance within the model equations does not change, and the dynamics are governed by Equations ({\ref{W eqn1}}) and ({\ref{A eqn1}}).
    \item Region II: As antibodies continue to bind free antigen, the number of available antigens decreases until they are almost saturated with antibody. As a result. the relative size of the terms in the model equations change. In Equations ({\ref{W eqn1}}) and ({\ref{A eqn1}}), the term $|1-W-A|$ is no longer of $\mathcal{O}(1)$. As a result, the dynamics of $W$ and $A$ are fundamentally different in Region II.
\end{itemize}
To summarise, Equations (\ref{beta neq 1/2 sol}) and (\ref{beta=1/2 sol}) are valid provided the dynamics lie within Region I. Eventually, as antibodies continue to bind, the system enters Region II and a new solution is needed which we outline below. A visual depiction of Regions I and II together with the inner and outer regions is presented in Figure {\ref{fig: regionI/II }}.
\subsubsection{$\beta = \text{ord}(1)$ Region II Method}\label{sec: Region II}
Here we construct solutions that remain valid in Region II where $W + A \approx 1$ for $\beta = \text{ord}(1)$. 
Recall that in the outer region
\begin{align}
    \frac{\mathrm{d}W}{\mathrm{d}\tau} &= 2\alpha_1(1-W-A^-(W))(\beta -W) - (W-A^-(W)),  \label{W region 2}
\end{align}
with $A^-(W)$ given by Equation (\ref{sol branches}).
Since we are interested in Region II where antigens are saturated with bivalently bound antibody (i.e. $W=A\approx 1/2)$, let 
\begin{align}
    & \zeta = 1 - 2W, \label{zeta def}\\
    & \lambda = \sqrt{\zeta^2 + 2\delta} -\delta = 1 - 2A^-(W), \label{lambda def}
\end{align}
where the final term neglects $\mathcal{O}(\delta^2)$ contributions. Additionally, we set
\begin{equation}
    \bar{\beta} = 2\beta -1,
\end{equation}
so that it is simpler to consider the cases $\beta<1/2$, $\beta=1/2$ and $\beta > 1/2$ (i.e. $\bar{\beta}<0$, $\bar{\beta}=0$ and $\bar{\beta}>0$).  Substituting the above expressions into Equation (\ref{W region 2}) we deduce that the system dynamics evolve as follows
\begin{equation}
    \frac{\mathrm{d} \zeta}{\mathrm{d} \tau} = (\lambda - \zeta) -\alpha_1(\zeta + \lambda)(\bar{\beta} + \zeta)  \eqqcolon F(\zeta), \label{F ode}
\end{equation}
with effective initial condition
\begin{equation}
    \zeta(\tau^*) = 1 - 4\alpha_1\beta\tau^*, \label{zeta init}
\end{equation}
for any $\tau^*$ satisfying $\delta\ll \tau^* \ll 1$, noting the behaviour of $W$ and $A$ in this temporal region from Equations (\ref{W inner sol})-(\ref{A qss eqn}), so that the initial condition is imposed after the initial transients. \par
We are interested in the long-time steady state behaviour of Equation (\ref{F ode}). Analysing $F(\zeta)=0$ further:
\begin{equation}
    F(\zeta) =  (\sqrt{\zeta^2 + 2\delta} - \delta - \zeta) - \alpha_1(\sqrt{\zeta^2 + 2\delta} -\delta + \zeta)(\bar{\beta} + \zeta)=0. \label{F expans}
\end{equation}
Noting
\begin{equation}
    a-b =\frac{a^2-b^2}{a+b},
\end{equation}
Equation (\ref{F expans}) becomes
\begin{align}
    F(\zeta) &= \frac{2\delta(1+ \sqrt{\zeta^2 + 2\delta} +\delta/2)}{\sqrt{\zeta^2 + 2\delta} -\delta + \zeta} - \alpha_1(\sqrt{\zeta^2 + 2\delta} -\delta + \zeta)(\bar{\beta}+\zeta), \\
    &=\frac{\alpha_1}{\sqrt{\zeta^2 + 2\delta} -\delta + \zeta}\biggl[ \frac{2\delta(1+ \sqrt{\zeta^2 + 2\delta} +\delta/2)}{\alpha_1} - (\sqrt{\zeta^2 + 2\delta} -\delta + \zeta)^2(\bar{\beta} + \zeta)\biggr]=0
\end{align}
So, $F(\zeta)=0$ has a unique root, which we will denote $\zeta^*$, when
\begin{equation}
    \frac{2\delta}{\alpha_1} = \frac{(\sqrt{\zeta^2 + 2\delta} -\delta + \zeta)^2(\bar{\beta} + \zeta)}{1+ \sqrt{\zeta^2 + 2\delta} +\delta/2} \eqqcolon G(\bar{\beta}, \delta, \zeta). \label{G eqn}
\end{equation}
For a proof that $F(\zeta)=0$ has a unique root $\zeta=\zeta^*$ see Appendix {\ref{f proof}}, and note that the root has $\bar{\beta} + \zeta^*>0$ since $2\delta/\alpha_1>0$. \par The sign of $\zeta^*$ is of interest because from Equation (\ref{zeta def}), we have $\zeta=1-2W$ so a value of $\zeta^*<0$ gives that $W>1/2$. Noting that $W=1/2$ corresponds with all antibodies being bivalently bound ($W=1$ similarly means all antibodies are monovalently bound), then $\zeta^*<0$ corresponds to a parameter regime that enables monovalent binding, with $|\zeta^*|$ correlating with the number of monovalently bound antibodies. \par
\begin{figure}[h!]\captionsetup[subfigure]{font=normal}
\captionsetup{width=1\textwidth}
\begin{tabular}{cccc}
\subfloat[$\bar{\beta} > 0 $]{\includegraphics[width=0.5\linewidth]{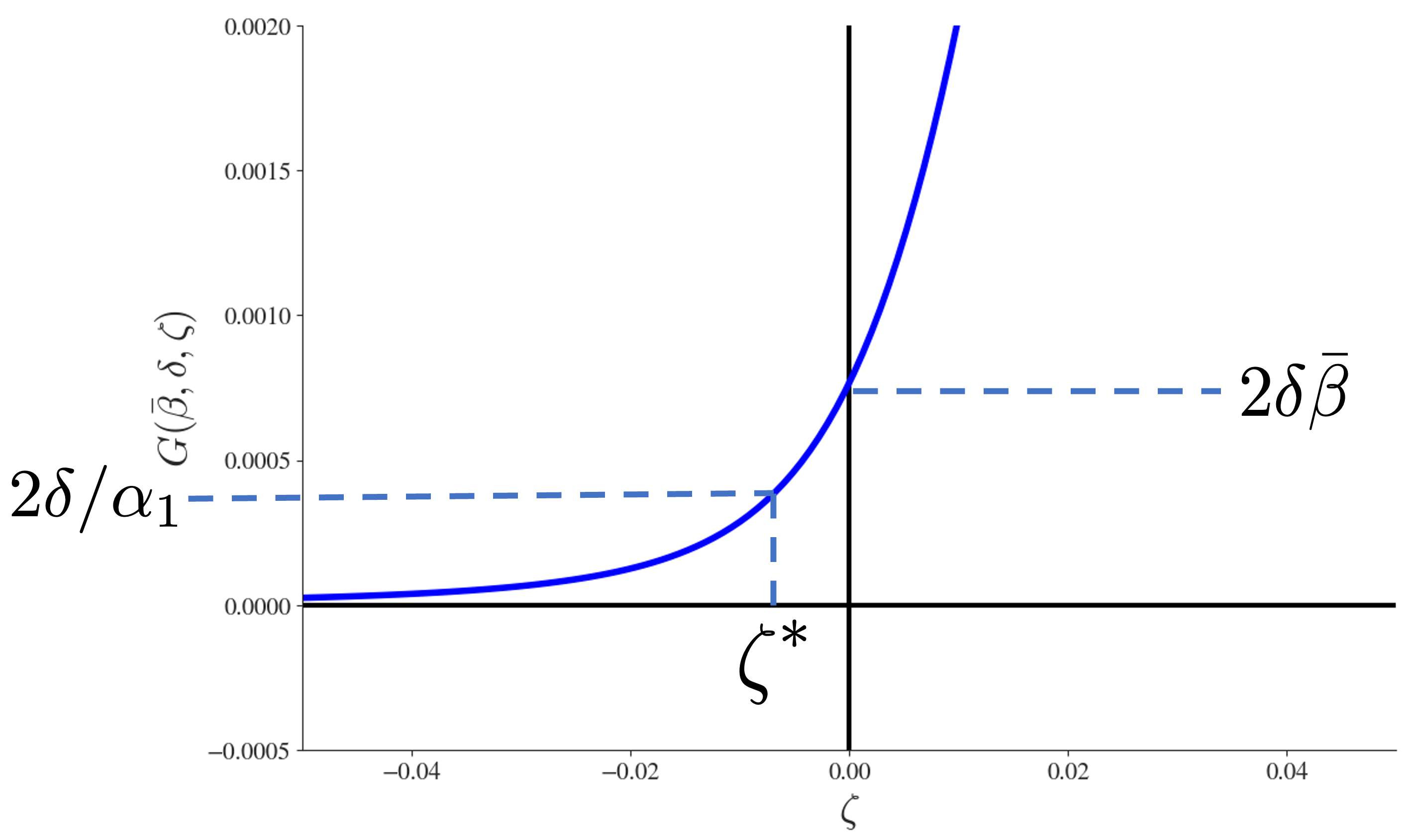}\label{G_beta_4}} &
\subfloat[$\bar{\beta} = 0$] {\includegraphics[width=0.5\linewidth]{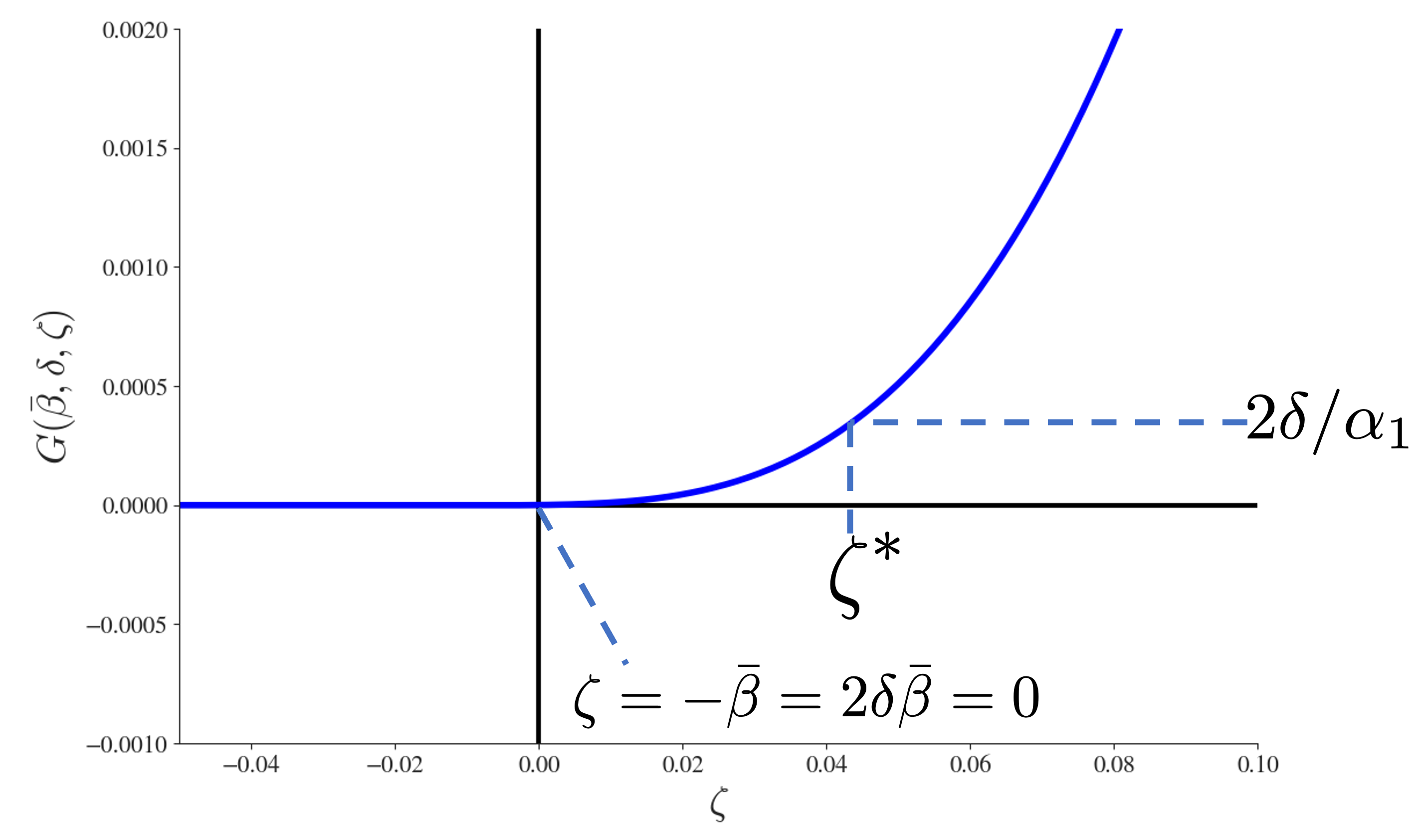}\label{G_beta_0}} \\
\end{tabular}
\subfloat[$\bar{\beta} < 0 $]{\includegraphics[width=0.5\linewidth]{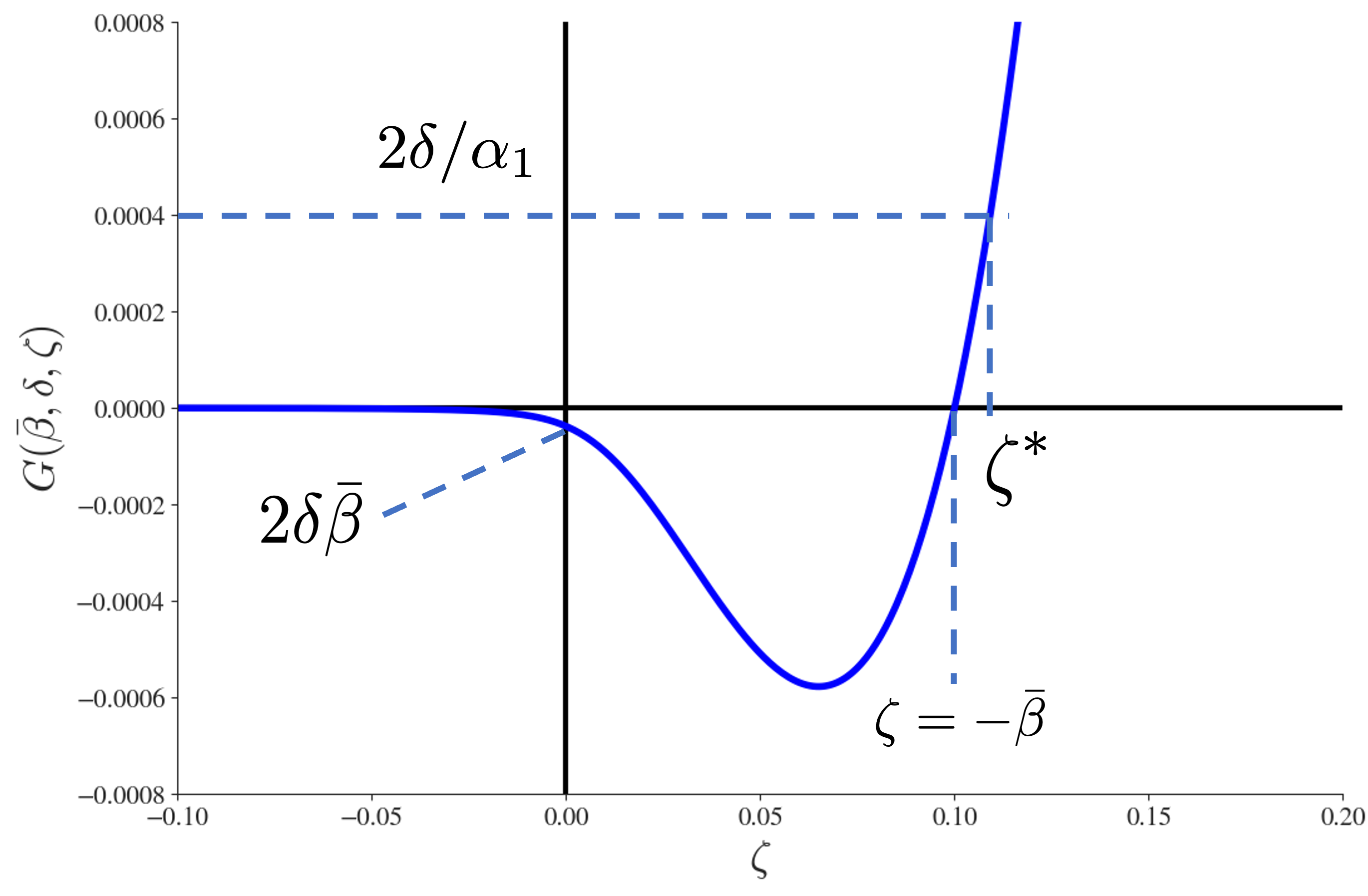}\label{G_beta_-4}} 
\centering
\caption{Plots of $G(\bar{\beta}, \delta, \zeta)$ as given by Equation (\ref{G eqn}) as $\zeta$ varies for different values of $\bar{\beta}$: (a) $\bar{\beta} >0$: there are more antigens than antibody, (b) $\bar{\beta}=0$: antibodies and antigens are in equal number, (c) $\bar{\beta} <0$: there are more antibodies than antigen.}
\label{fig: G figs}
\end{figure}

Figure \ref{fig: G figs} shows example plots of $G(\bar{\beta}, \delta, \zeta)$ and the corresponding roots $\zeta^*$ for different values of $\bar{\beta}$. We note that $\zeta^*>0$ in Figures \ref{G_beta_-4} and \ref{G_beta_0} while $\zeta^*<0$ in Figure \ref{G_beta_4}. In particular, as illustrated in Figure \ref{fig: G figs}, $\zeta^*<0$ if and only if 
\begin{equation}
     \alpha_1\bar{\beta} > 1. \label{mono relation}
\end{equation}
Therefore, a necessary and sufficient condition for antibodies to be monovalently bound is $\alpha_1\bar{\beta} >1$. We conclude that if there are few antibodies (low value of $\bar{\beta}$) then antibodies can still be monovalently bound on the cell surface if the monovalent binding rate is sufficiently large (high value of $\alpha_1$) and vice versa. Interestingly, Equation (\ref{mono relation}) is independent of $\alpha_2$, the rate of the bivalent reaction.
\par
Finding $\zeta^*$ gives the long-time behaviour of $W$ and $A$ within Region II, but it still remains to find a leading order estimate of $A_1$ in this regime. As $\tau \rightarrow \infty$, we have:
\begin{align}
    A_1 &= W-A\approx W - A^-(W) = \frac{1}{2}\biggl( \lambda - \zeta \biggr) = \frac{1}{2}\biggl( \frac{2\delta(1 - \sqrt{\zeta^2 + 2\delta} + \delta/2)}{\sqrt{\zeta^2+2\delta} -\delta + \zeta} \biggr)\\ &\rightarrow  \sqrt{\frac{\delta\alpha_1(1 - \sqrt{(\zeta^{*})^2 + 2\delta} + \delta/2)}{2}\biggl(\bar{\beta} + \zeta^*\biggr)}. \label{A1 region 2}
\end{align}
Upon finding the unique root to $F(\zeta)=0$ given by $\zeta=\zeta^*$, a leading order approximation to $A_1$ can be found by substituting this root into Equation (\ref{A1 region 2}). This provides a framework to obtain $W$ and $A_1$ for the cases $\beta < 1/2$, $\beta=1/2$ and $\beta > 1/2$. We will consider the cases when $\beta = \text{ord}(1/\epsilon)$ and $\beta = \text{ord}(1/\epsilon^2)$ separately in later sections.\par
Due to their impact on mAb treatment potency and efficacy, we are interested in the long-time behaviour of quantities such as antigen occupancy and the number of bound and monovalently bound antibodies \mbox{\citep{Mazor2016, Junker2021}}. From our asymptotic analysis, these values correspond to the Region II large time asymptotes of the following quantities
\begin{align}
    W &= A_1 + A_2  = \frac{1- \zeta}{2}, \label{bound ab}\\
    A_1 &= W - A_2 = \frac{\lambda - \zeta}{2}, \label{mono ab}\\
    W + A_2 &= A_1 + 2A_2 = 1 - \frac{\lambda + \zeta}{2}. \label{RO}
\end{align}
We will estimate these quantities for the different magnitudes of $\beta$ in the following sections.
\subsection{Long Time Asymptotes of Region II Solution for $\beta=\text{ord}(1)$}
In this section we summarise in Table \ref{Summary zeta table} below the values of the unique root $\zeta^*$ and corresponding values of $W$, $A_1$ and $A_1 + 2A_2$ for the different cases when $\beta=\text{ord}(1)$. All details on the calculations of the respective $\zeta^*$ can be found in Appendix \ref{zeta appendix}.
\begin{table}[h!]
\centering
\caption{Asymptotic approximations for the large time values of the unique root to Equation (\ref{F expans}), $\zeta^*$, antigen occupancy, $A_1 + 2A_2,$ and total, $W$, and monovalently bound, $A_1$, antibody numbers. Recall that $\beta= A_{\text{tot}}/r_{\text{tot}}$ is the ratio of total antibody to receptor within the system (with $\bar{\beta} = \beta-1/2$), $\alpha_1$ and $\alpha_2$ (see Equation (\ref{nondim express})) are the nondimensional monovalent and bivalent binding rates respectively and $\delta = 2\epsilon^2/\hat{\alpha}_2$ where $\hat{\alpha}_2= \epsilon^2 \alpha_2=\text{ord}(1)$. Details on the calculations to obtain these expressions can be found in Appendix \ref{zeta appendix}.}\label{Summary zeta table}  
\begin{tabular}[\textwidth]{@{}lllll@{}} 
\toprule
Value of $\bar{\beta}$ & $\zeta^*$ estimate& \textbf{$A_1 + 2A_2$} & \textbf{$W$} & \textbf{$A_1$} \\
\hline
$\bar{\beta}<0$ &  $|\bar{\beta}| + \frac{\delta(1 + |\bar{\beta}|)}{2|\bar{\beta}|^2\alpha_1}$& $1-|\bar{\beta}|$ & $\frac{1}{2}\biggl(1 - |\bar{\beta}|\biggr)$ & $\frac{\delta(1 + |\bar{\beta}|)}{2|\bar{\beta}|}$ \\ 
\hline
$\bar{\beta} > 0$ & $-\sqrt{\frac{\delta}{2\alpha_1\bar{\beta}}}(\alpha_1\bar{\beta} - 1)$&$\approx 1 $ & $ \frac{1}{2}\biggl(1 + \sqrt{\frac{\delta}{2\alpha_1\bar{\beta}}}(\alpha_1(\bar{\beta}) - 1)\biggr)$ & $\sqrt{\frac{\delta \alpha_1(\bar{\beta})}{2}}$ \\ 
\hline
$\bar{\beta}=0 $ & $\biggl( \frac{\delta}{2\alpha_1}\biggr)^{1/3}$& $\approx 1 $ & $\frac{1}{2}\biggl(1 -  \biggl( \frac{\delta}{2\alpha_1}\biggr)^{1/3}\biggr)$ &  $\biggl(\frac{\delta^2\alpha_1}{4}\biggr)^{1/3}$\\ 
\botrule
\end{tabular}
\end{table}\par

\begin{figure}[h!]\captionsetup[subfigure]{font=normal}
\captionsetup{width=1\textwidth}
\begin{tabular}{cccc}
\subfloat[$\beta < 1/2$]{\includegraphics[width=0.5\linewidth]{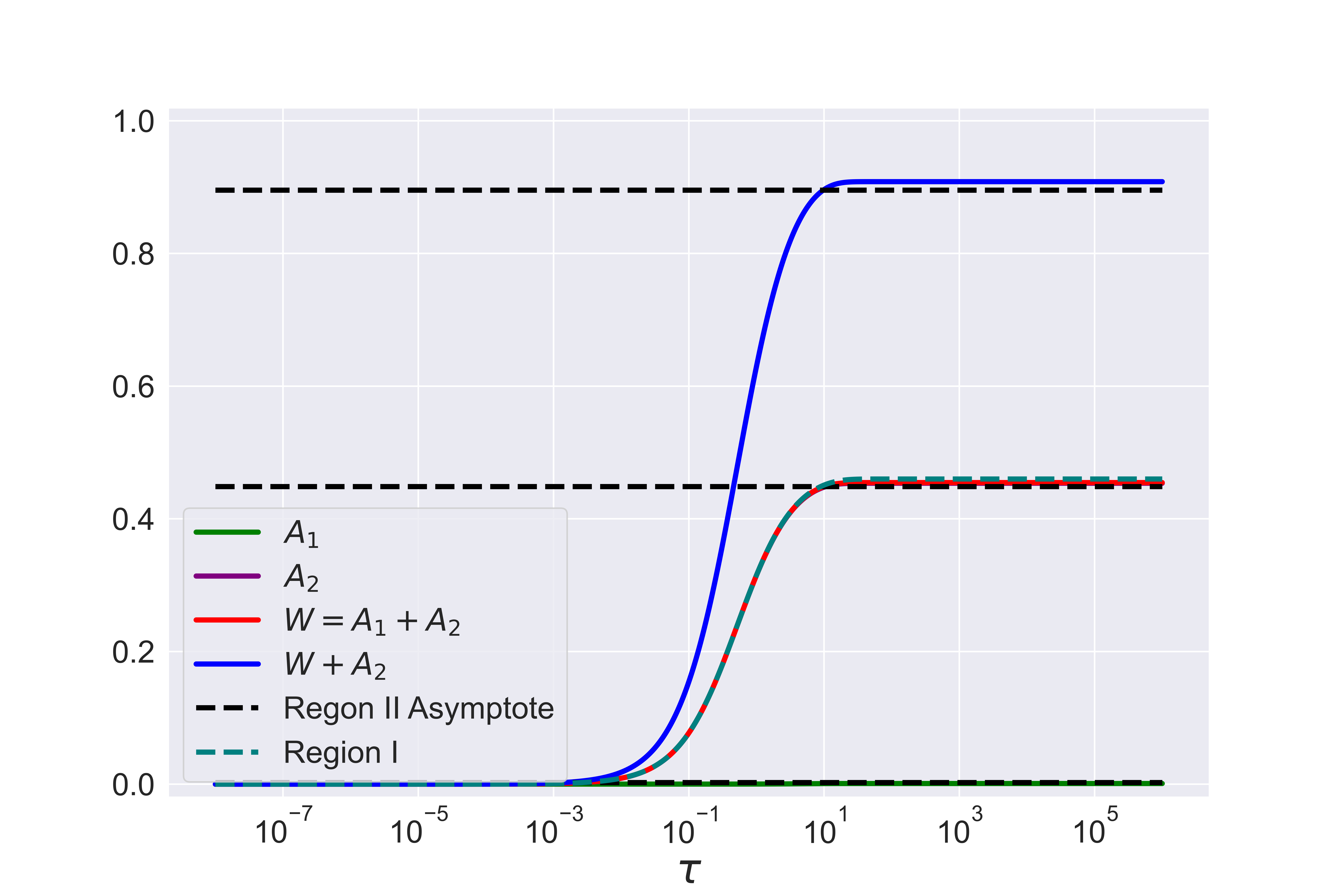}\label{sim_0.25}} &
\subfloat[$\beta = 1/2$] {\includegraphics[width=0.5\linewidth]{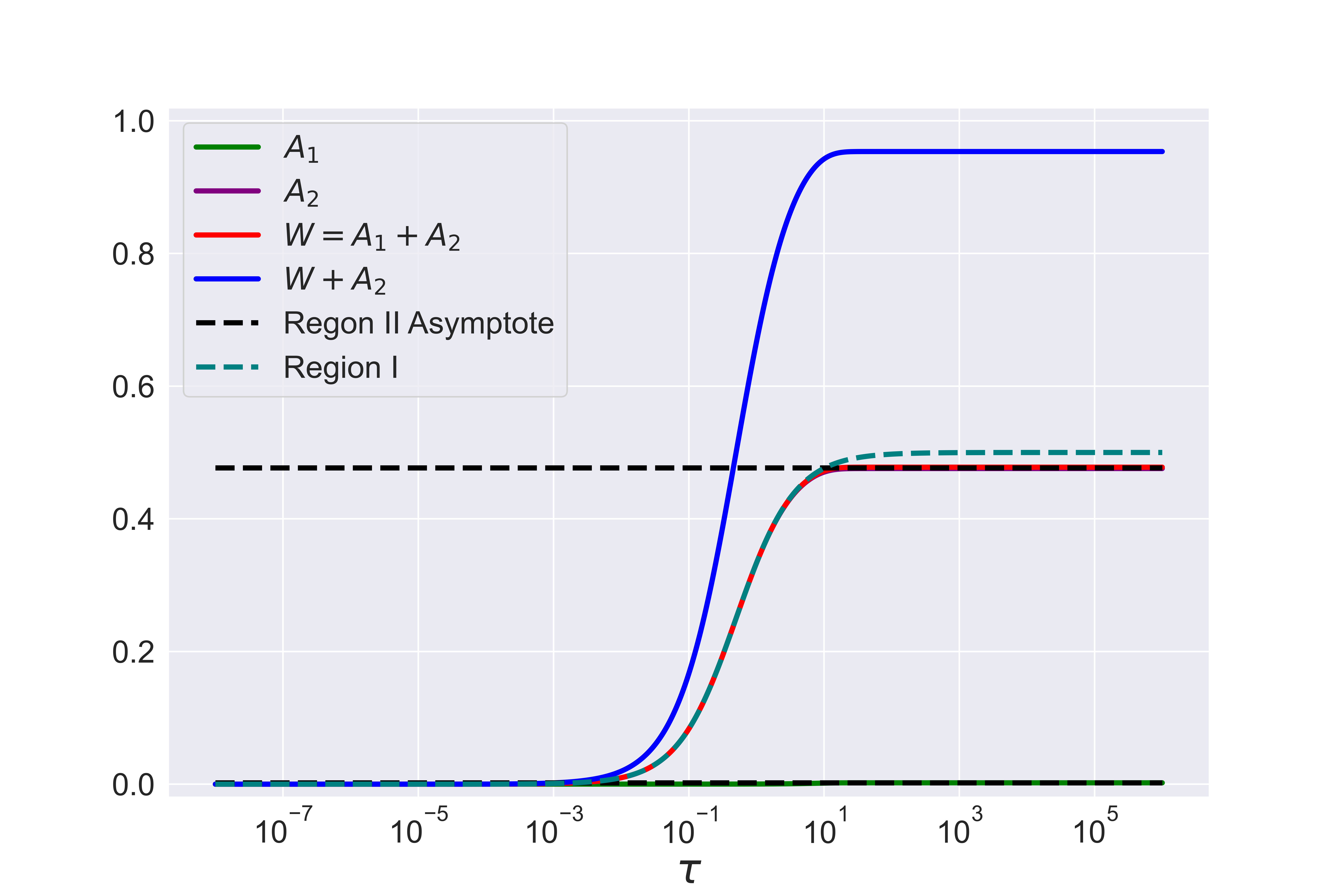}\label{sim_0.5}} \\
\end{tabular}
\subfloat[$\beta > 1/2 $]{\includegraphics[width=0.5\linewidth]{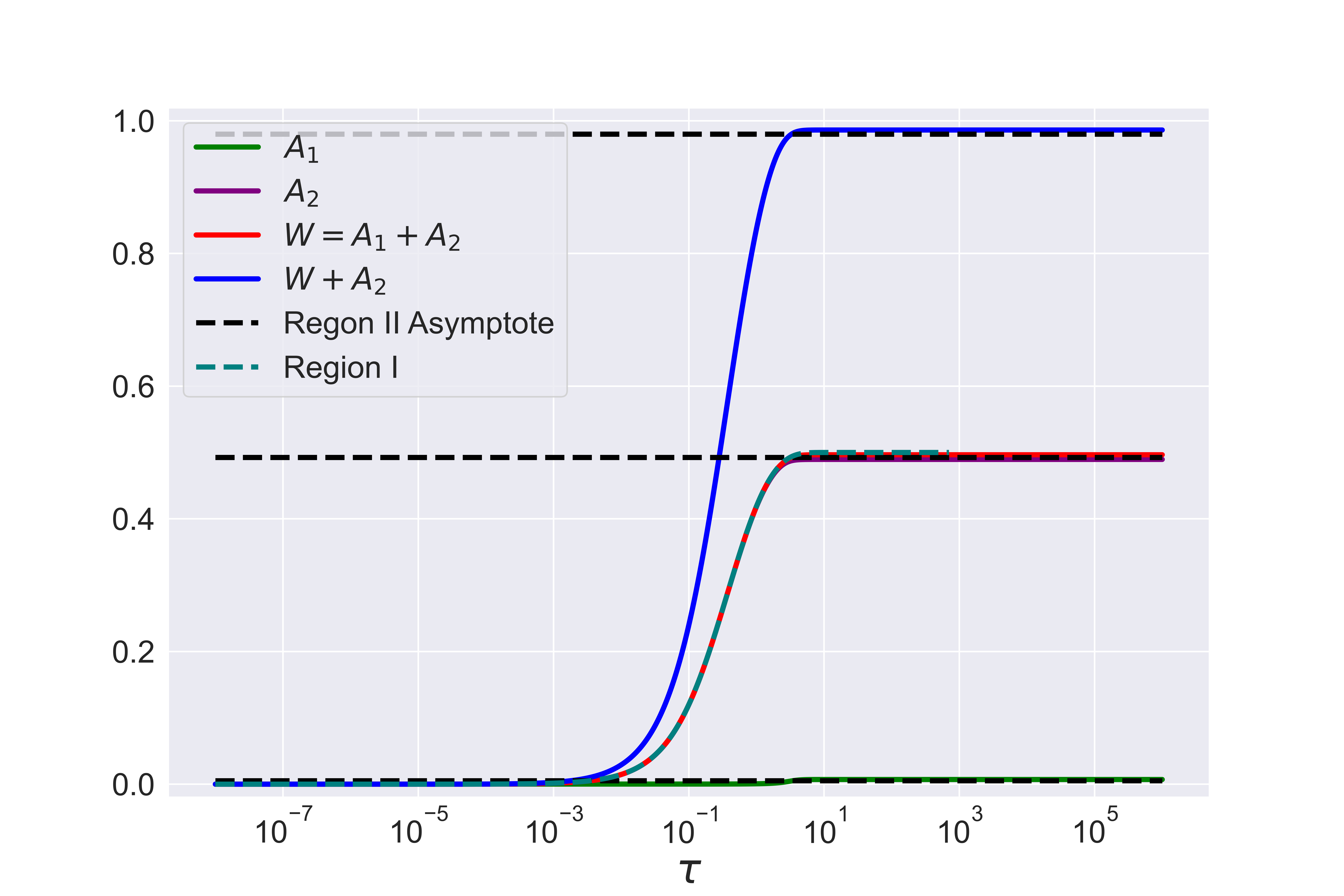}\label{sim_0.75}} 
\centering
\caption{Model Dynamics from Equations (\ref{nondim1})-(\ref{nondim ics}) for the three cases when $\beta = \text{ord}(1)$; (a): $\beta=0.46<1/2$, (b): $\beta=1/2$, (c): $\beta =0.75>1/2$ with $\alpha_1=1$, $\alpha_2=10^4$. Asymptotic approximations to the model dynamics in Region I and the long-time asymptotes in Region II solutions respectively correspond to the teal and black dashed lines.}
\label{fig: ord 1 dynamics}
\end{figure}

Table \ref{Summary zeta table} shows that antigens are fully occupied at leading order for $\beta>1/2$ and $\beta=1/2$ ($\bar{\beta}>0$ and $\bar{\beta}=0$ respectively.) This is to be expected as for these values of $\beta$ there are enough antibodies to saturate all antigens, in contrast to when $\beta <1/2$. Figure \ref{fig: ord 1 dynamics} shows that the results of the asymptotic analysis for $\beta = \text{ord}(1)$ are in good agreement with the numerical solution of the full system. 
\subsection{Perturbation Analysis for $\beta = \text{ord}(\epsilon^{-1})$}
For $\beta=\text{ord}(1/\epsilon)$, there are many more antibodies than antigens and the system dynamics are governed by the following equations:
\begin{align}
    \frac{\mathrm{d}W}{\mathrm{d}\tau} &= 2\alpha_1(1-W-A)\biggl(\frac{\hat{\beta}}{\epsilon}-W\biggr) - (W-A), \label{W 1/epsilon}\\
    \frac{\mathrm{d}A}{\mathrm{d}\tau} &= \frac{2}{\delta}\biggl[(W-A)(1-W-A) -\delta A\biggr], \label{A 1/epsilon} 
\end{align}
In order to proceed, as for the $\beta=\text{ord}(1)$ case, we must first show that $A$ is at quasi-steady state away from initial transients so that we can utilise the framework presented in Section {\ref{sec: Region II}} for Region II. As before, at sufficiently small $\tau$, we have that $W,A \ll 1$ since $W(0)=A(0)=0$. Equation ({\ref{W 1/epsilon}}) then becomes
\begin{equation}
    \frac{\mathrm{d}W}{\mathrm{d}\tau} = 2\alpha_1\frac{\hat{\beta}}{\epsilon}, \label{W fast}
\end{equation}
and so $W>0$ for $\tau>0$. While $W,A \ll 1$, Equation ({\ref{A 1/epsilon}}) becomes 
\begin{equation}
    \frac{\mathrm{d}A}{\mathrm{d}\tau} = \frac{2}{\delta}\biggl[W - A \biggr],
\end{equation}
which, following the same method as for $\beta=\text{ord}(1)$, has the solution
\begin{equation}
    A = W(\tau)(1 - e^{-\frac{2}{\delta}\tau}) + \mathcal{O}(\delta). \label{1/e qss}
\end{equation}
Therefore, $A$ reaches the quasi-steady state $W(\tau)$ for $\tau \gg \delta/2$. The condition for Equation ({\ref{1/e qss}}) to be valid is that $W,A \ll 1$. It can be shown from Equation ({\ref{W fast}}) that $W \centernot{\ll} 1$ once $\tau\approx \text{ord}(\delta^{1/2})$. Therefore, the quasi-steady state in Equation ({\ref{1/e qss}}) is valid provided $\delta \ll \tau \ll \delta^{1/2}$. More generally, $A$ reaches the quasi-steady state given by the branch $A=A^-(W)$ for $\tau \gg \delta$ by the same reasoning presented for $\beta=\text{ord}(1)$. Hence, the Region II approximation has the same structure as previously. We now exploit this structure to determine the long time behaviour of the system in Region II.
\subsubsection{Long Time Asymptote of Region II Solution for $\beta=\text{ord}(1/\epsilon)$}
\begin{figure}[h!]
    \centering
    \includegraphics[width=1\textwidth]{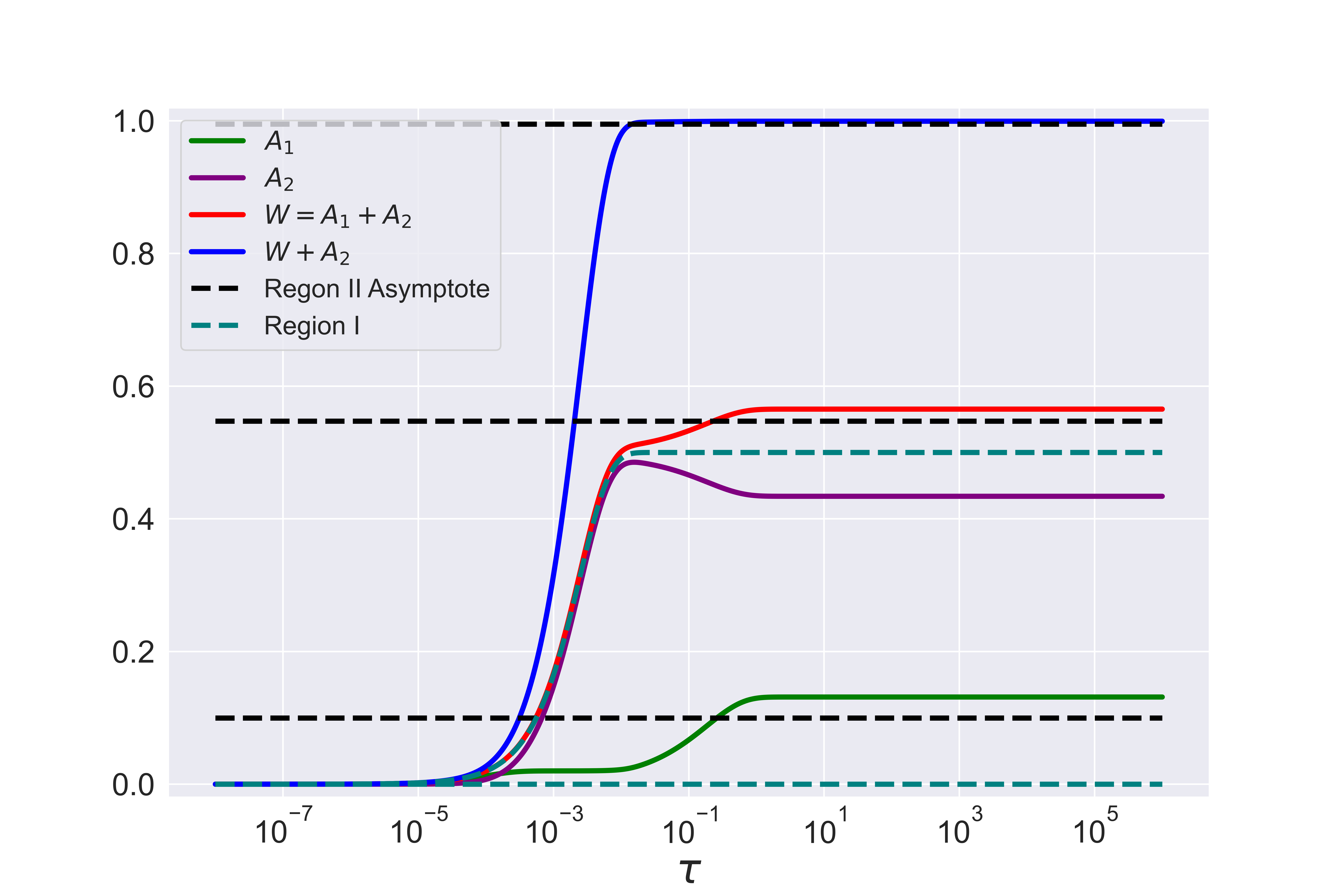}
    \caption{Result of asymptotic analysis of Equations (\ref{W 1/epsilon}) and (\ref{A 1/epsilon}) for $\beta = \text{ord}(1/\epsilon)$ The inner solution is given by Equation (\ref{beta neq 1/2 sol}) and denoted by the dashed teal line. The Region II asymptotes are given by Equations (\ref{1/epsilon II W}) and (\ref{1/epsilon II A}) and are denoted with dashed black lines.}
    \label{fig:1e2 summary}
\end{figure}

As $W$ and $A$ both grow close to 1/2, we will eventually enter Region II  where $|1-W-A| = \mathcal{O}(\epsilon^2)$. We then can find the unique root to $F(\zeta)$  within the physical regime of $\zeta$ and, in so doing, find the long-time asymptotes of $W$ and $A_1$ in this regime. Following Appendix \ref{sec: zeta beta>1/2}, we obtain
\begin{equation}
    \zeta^* = -\sqrt{\frac{\delta}{2\alpha_1\bar{\beta}}}(\alpha_1\bar{\beta} - 1) \approx -\sqrt{\frac{\delta \alpha_1 \bar{\beta}}{2}}, \label{zeta^* 1/epsilon}
\end{equation}
so that
\begin{equation}
    W \rightarrow \frac{1}{2}\biggl(1 + \sqrt{\frac{\delta \alpha_1 \bar{\beta}}{2}} \biggr) \text{   as   } \tau \rightarrow \infty, \label{1/epsilon II W}
\end{equation}
and
\begin{equation}
    A_1 \rightarrow \sqrt{\frac{\delta \alpha_1}{2}\biggl(\bar{\beta} - \sqrt{\frac{\delta \alpha_1 \bar{\beta}}{2}}\biggr)} \approx \sqrt{\frac{\delta\alpha_1 \bar{\beta}}{2}}  \text{   as   } \tau \rightarrow \infty \label{1/epsilon II A},
\end{equation}
(where we have used the fact that $\delta \ll \bar{\beta}$). 
Figure \ref{fig:1e2 summary} shows that the results of the above asymptotic analysis are in good agreement with the numerical solutions of the full system. \par
From Equations (\ref{1/epsilon II W}) and (\ref{1/epsilon II A}), it is clear that the  total number of bound antibodies and the number of monovalently bound antibodies increases with $|\zeta^*|$, provided $|\zeta^*| \leq 1$. Rewriting $|\zeta^*|$ in dimensional units gives
\begin{equation}
    |\zeta^*| = \sqrt{\frac{A_{\text{init}}k_{\text{on}}}{r_{\text{tot}}k_2}} = \text{ord}(\epsilon^{1/2}).
\end{equation}
It follows that for this value of $\beta$, quantities such as the number of bound antibodies, $W$, can be increased by, for example, decreasing the the bivalent binding rate, $k_2$. Considering antigen occupancy, we have that
\begin{equation}
    2W - A_1 \approx 1,
\end{equation}
i.e. receptors are saturated with antibody. 
\subsection{Perturbation Analysis for $\beta = \text{ord}(1/\epsilon^2)$}
When $\beta = \text{ord}(\epsilon^{-2})$, we have
\begin{align}
    \frac{\mathrm{d}W}{\mathrm{d}\tau} &= 2\alpha_1(1-W-A)\biggl(\frac{\hat{\beta}}{\epsilon^2}-W\biggr) - (W-A), \label{W 1/epsilon^2} \\
    \frac{\mathrm{d}A}{\mathrm{d}\tau} &= \frac{\hat{\alpha_2}}{\epsilon^2}\biggl[(W-A)(1-W-A) -\delta A\biggr], \label{A 1/epsilon^2}
\end{align}
with
\begin{equation}
    W(0)=A(0)=0.
\end{equation}
\subsubsection{Region I Solution for $\beta = \text{ord}(1/\epsilon^2)$}\label{sec: 1/epsilon^2}
As with the case of $\beta = \text{ord}(\epsilon^{-1})$, we have a singular perturbation problem with a boundary layer at $\tau=0$ in which the dynamics evolve on a short timescale. To capture this, we rescale
\begin{equation}
    \tau = \frac{\epsilon^2\bar{\tau}}{\hat{\alpha_2}}. \label{fast time 1}
\end{equation}
Substituting Equation (\ref{fast time 1}) into Equations (\ref{W 1/epsilon^2}) and (\ref{A 1/epsilon^2}) supplies
\begin{align}
    \frac{\mathrm{d}W}{\mathrm{d}\bar{\tau}} &= \frac{2\alpha_1\hat{\beta}}{\hat{\alpha_2}}(1-W-A) - \frac{\delta}{2}(W-A), \label{W inner 1} \\
    \frac{\mathrm{d}A}{\mathrm{d}\bar{\tau}} &= (W-A)(1-W-A) - \delta A. \label{A inner 1}
\end{align}
We simplify the analysis of Equations (\ref{W inner 1}) and (\ref{A inner 1}), by defining  $p=\zeta + \lambda = 2(1-W-A)$ and $q = \lambda - \zeta = 2(W-A)$. Provided $|1-W-A|\gg \epsilon^2$ (which is true initially as $W(0)=A(0)=0$), recasting Equations (\ref{W inner 1}) and (\ref{A inner 1}) in terms of $p$ and $q$ supplies at leading order
\begin{align}
    \frac{\mathrm{d}p}{\mathrm{d}\bar{\tau}} &= p\biggl(-\mu - \frac{q}{2}\biggr), \label{p eqn} \\
    \frac{\mathrm{d}q}{\mathrm{d}\bar{\tau}} &= p\biggl(\mu - \frac{q}{2}\biggr), \label{q eqn}
\end{align}
where $\mu=2\alpha_1\hat{\beta}/\hat{\alpha_2}$ and $p(0) = 2, q(0)=0$. Using Equations  { (\ref{p eqn})} and ({\ref{q eqn}}), it can be shown that $q$ attains a steady state $q^*$ as $\tau \rightarrow \infty$ (for details, see Appendix {\ref{region I beta big}}), so that
\begin{equation}
    p \rightarrow 0 \text{   as   } \tau \rightarrow \infty. \label{inner asymptote 1}
\end{equation}
From which we deduce
\begin{equation}
    A+W \rightarrow 1 \text{   as   } \tau \rightarrow \infty,
\end{equation}
so all antigens become bound with antibody in Region I. We also note for later use that $p>0$ as it approaches the limit of Equation (\ref{inner asymptote 1}) since $p(0)=2, q(0)=0$.
\subsubsection{Region II Solution for $\beta = \text{ord}(1/\epsilon^2)$}
The number of free antigens becomes very small, and eventually $|1-W-A| = \mathcal{O}(\epsilon^2)$, transitioning the system into Region II. For $\beta = \text{ord}(1/\epsilon^2)$, in contrast to $\beta=\text{ord}(1)$ or $\beta = \text{ord}(1/\epsilon)$, the inner region corresponds to Region I and the outer region corresponds to Region II. Furthermore, the largest terms in both Equation (\ref{W 1/epsilon^2}) and (\ref{A 1/epsilon^2}) are $\mathcal{O}(1/\epsilon^{2})$ so we can no longer assume $A$ is at quasi-steady state.\par
As we enter the outer region/Region II, we have $|1-W-A| = \mathcal{O}(\epsilon^2)$. Thus,
\begin{equation}
    0 < p = \zeta + \lambda = 2(1-W-A) \approx \mathcal{O}(\epsilon^2), \label{zeta lambda approx}
\end{equation}
and 
\begin{equation}
    q = \lambda- \zeta \rightarrow q^* >0, \label{q approx}
\end{equation}
therefore $\lambda  >0 > \zeta$. Rewriting Equations (\ref{W 1/epsilon^2}) and (\ref{A 1/epsilon^2}) in terms of $\zeta$ and $\lambda$ we have 
\begin{align}
    \frac{\mathrm{d}\zeta}{\mathrm{d} \tau} &= -\frac{\mu}{\epsilon^2}(\zeta + \lambda) + \frac{(\lambda - \zeta)}{\hat{\alpha_2}}, \label{zeta full1}\\
    \frac{\mathrm{d} \lambda}{\mathrm{d} \tau} &= -\frac{1}{2\epsilon^2}(\lambda-\zeta)(\zeta+\lambda) + \frac{2}{\hat{\alpha_2}}(1-\lambda). \label{lambda full1}
\end{align}
Using Equations {(\ref{zeta full1})} and {(\ref{lambda full1})}, it can be shown (for details, see Appendix {\ref{region II beta big}}) that within Region II
\begin{align}
    \zeta &\approx \frac{\mu}{2}\biggl(1 - \sqrt{1 + \frac{4}{\mu}}\biggr) \text{   as   } \tau \rightarrow \infty,\label{zeta sol1} \\
    \lambda &\approx - \frac{\mu}{2}\biggl(1 - \sqrt{1 + \frac{4}{\mu}}\biggr) + \epsilon^2\frac{2}{\hat{\alpha}_2 \mu} \text{   as   } \tau \rightarrow \infty.\label{lambda sol1}
\end{align}
Simplifying Equations (\ref{zeta sol1}) and (\ref{lambda sol1}) by Taylor expanding the square root and substituting the resulting expressions into Equations (\ref{bound ab})-(\ref{RO}) gives
\begin{align}
    W & \approx 1 - \frac{1}{2\mu}, \label{bound ab approx}\\
    A_1 &\approx 1 - \frac{1}{\mu} + \frac{\epsilon^2}{\hat{\alpha}_2\mu},  \label{mono ab approx}\\
    A_1 + 2A_2 & \approx 1 - \frac{\epsilon^2}{\hat{\alpha_2}\mu}.\label{RO approx}
\end{align}
From Equation (\ref{RO approx}) we see that at leading order antigens are saturated with antibody independent of the model parameters. Therefore, immune checkpoint inhibitors, whose efficacy relies on antigen occupancy, will be insensitive to changes in model parameters in this region of parameter space. From Equations (\ref{bound ab approx}) and (\ref{mono ab approx}), both the total number of bound and number of monovalently bound antibodies only depend on the ratio $\mu= 2\alpha_1 \hat{\beta}/ \hat{\alpha}_2$, with these quantities increasing as the value of $\mu$ increases. In dimensional units
\begin{equation}
    \mu = \frac{2A_{\text{init}}k_{\text{on}}}{r_{\text{tot}}k_2}.
\end{equation}
Pertinent for the mechanism of ADCC, it follows that the number of bound antibodies can be increased by, for example, increasing the monovalent binding rate, $k_{\text{on}}$, or decreasing $k_2$, the rate at which the second arm bids. Interestingly, there is no dependence on the off- rate $k_{\text{off}}$ in $\mu$ and, as a result, the bound antibody number. This is consistent with the global parameter sensitivity analysis in \cite{Heirene2024} but highlights the quantitative parameter dependencies.\par
\begin{figure}
    \centering
    \includegraphics[width = 1\textwidth]{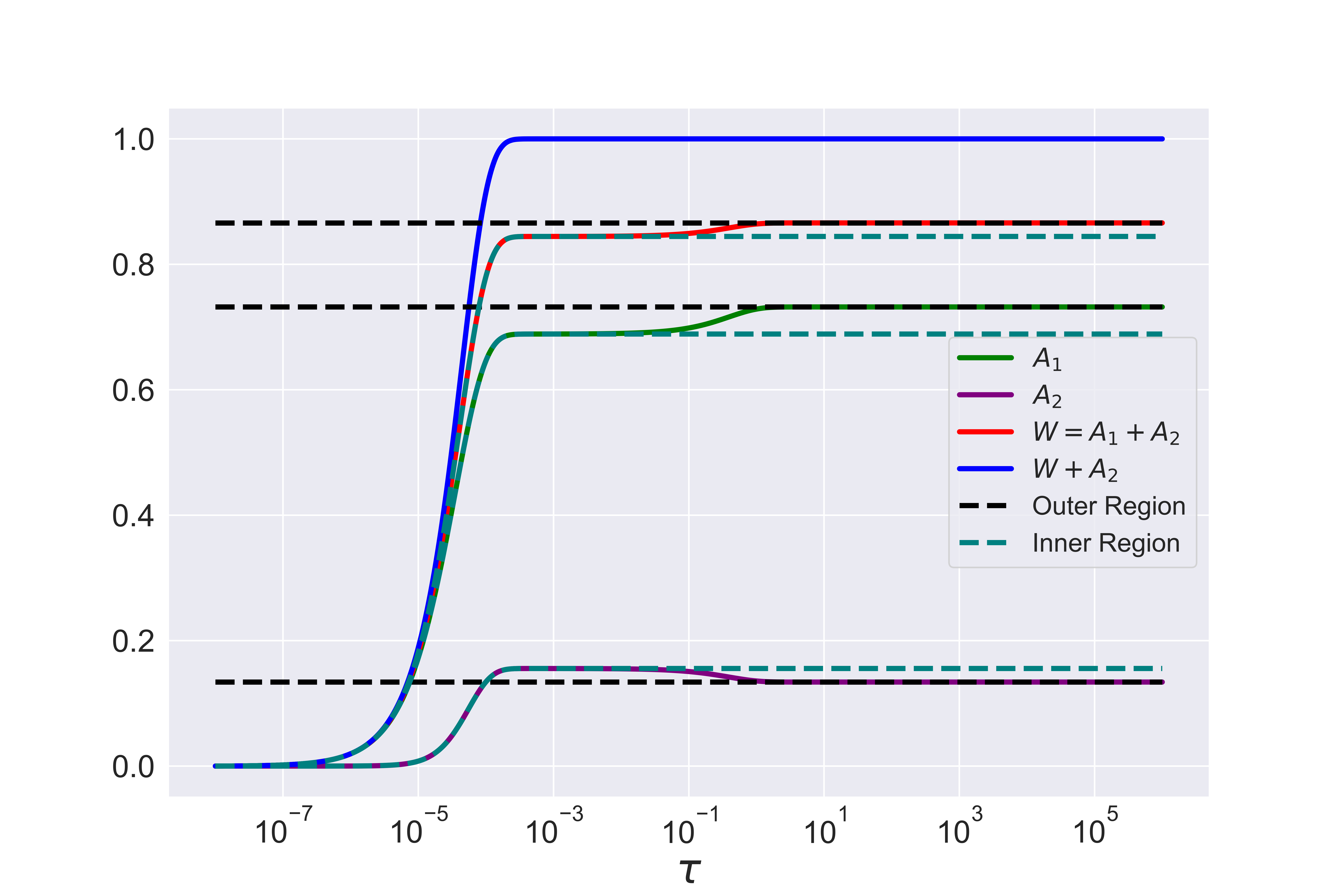}
    \caption{Result of asymptotic analysis of Equations (\ref{W 1/epsilon^2}) and (\ref{A 1/epsilon^2}) for $\beta = \text{ord}(1/\epsilon^2)$. Region I solutions are obtained by numerically solving Equations ({\ref{p eqn}}) and ({\ref{q eqn}}).The Region II solution asymptotes for $W$ and $A$ are gained by substituting Equations (\ref{zeta sol1}) and (\ref{lambda sol1}) into Equations (\ref{zeta def}) and (\ref{lambda def}) and are denoted with dashed black lines. Similarly, $A_1 = W-A$.}
    \label{fig: 1e4 summary}
\end{figure}

Figure \ref{fig: 1e4 summary} shows that the results of the asymptotic analysis for Region II are in excellent agreement with the numerical solutions of the full system. The value the number of bound antibodies, $W$, increases to when $\beta=\text{ord}(1/\epsilon^2)$ is larger than the value attained for smaller values of $\beta$, particularly when $\beta=\text{ord}(1)$. Consequently, when $\beta=\text{ord}(1/\epsilon^2)$ there may be increased mAb therapeutic effects due to increased bound antibody.\par
\section{Discussion}\label{sec: discussion}
Ligand-receptor interactions play a crucial role in drug mechanism of action. Of particular importance are antibody interactions with their cognate antigens. MAbs are antibody-based immunotherapies whose therapeutic effect stems from the dynamics of antibody-antigen binding. In this work, we used asymptotic analysis to identify the key features and parameter dependencies of mAb binding target antigens on a tumour cell's surface for a variety of antibody concentrations.\par
This work was motivated by two research aims, which we consider in turn. First we sought to describe and quantify the complex temporal dynamics of bivalent antibody-antigen binding across antibody concentrations commonly seen within in vitro experiments. This was achieved by performing asymptotic analysis for different magnitudes of $\beta = A_{\text{tot}}/r_{\text{tot}}$.
While many ligand-receptor interactions are monovalent, the binding dynamics of mAbs are more complex due to the bivalency of the antibody. This complexity is introduced through a fast-cross-linking reaction on the cell surface assumed to be driven by the surface diffusion of target antigen as motivated by \cite{Sengers2016, Heirene2024}. \par
Within our analysis, we constructed when appropriate inner and outer solutions to the model equations. Initially, there is a fast transient after which the number of bivalently bound antibodies, $A$, reaches a quasi-steady state that depends on the total bound antibody number, $W$. After this fast transient, $W$ and $A$ grow to become $\text{ord}(1)$ and the dynamics are governed by the outer solution. Through our asymptotic analysis, for the case where antibody is more abundant than antigen, we identified two sub-regions within the outer region: Region I where free antigen is abundant to bind antibody and Region II where levels of free antigen are very small. Analogous results hold for the case where antigen numbers exceed ligand.  Within  Region II, the number of bivalently bound antibodies decreases and the number of monovalently bound antibodies increases. The increase in monovalently bound antibodies further increases the number of bound antibodies as shown in Figure {\ref{fig:1e2 summary}} where both the total number of bound antibodies and the number of monovalently bound antibodies increase. \par
An interesting conclusion can be drawn from the fact the dynamics separate into two regions when antibodies exceed antigens; only dissociation events occur when the number of free antigens becomes very small. This is not immediately apparent from Equations (\ref{W eqn}) and (\ref{A eqn}) because the dominant process within these equations is the binding of the second arm of the antibody. Therefore, an antibody successfully dissociates one of its binding arms away from a bound antigen only when the number of free antigens is small and the binding and dissociation terms are balanced.\par
The second aim of this work has been to use asymptotic analysis to quantify how quantities that correlate with the potency and efficacy of mAb treatment depend explicitly on model parameters. In particular, we are interested in the long-time values of antigen occupancy and the number of total bound and monovalently bound antibodies (given by $A_1 + 2A_2$, $W$ and $A_1$ respectively). In our previous work we performed a global parameter sensitivity analysis to explore the general parameter dependencies of these quantities \citep{Heirene2024}. In contrast, here, we used asymptotic analysis to derive explicit expressions for their long-time values for the four considered magnitudes of $\beta$. See Table {\ref{Summary table}} for a summary of these approximations.
\begin{table}[h!]
\centering
\caption{Asymptotic approximations for the large time values of antigen occupancy, $A_1 + 2A_2,$ and total, $W$, and monovalently bound, $A_1$, antibody numbers. Recall that $\beta= A_{\text{tot}}/r_{\text{tot}}$ is the ratio of total antibody to receptor within the system, $\alpha_1$ and $\alpha_2$ (see Equation (\ref{nondim express})) are the nondimensional monovalent and bivalent binding rates respectively and $\delta = 2\epsilon^2/\hat{\alpha}_2$ where $\hat{\alpha}_2= \epsilon^2 \alpha_2=\text{ord}(1)$.}\label{Summary table}  
\begin{tabular}{@{}llll@{}} 
\toprule
\textbf{Magnitude of $\beta$} & \textbf{$A_1 + 2A_2$} & \textbf{$W$} & \textbf{$A_1$} \\
\hline
$\text{ord}(\epsilon)$ & $2\epsilon \beta$ & $\epsilon \beta$ & $2\epsilon^3\beta / \hat{\alpha}_2$ \\ 
\hline
$\text{ord}(1), |\beta-1/2|\gg \delta $ & $1-|\beta - \frac{1}{2}|$ & $\frac{1}{2}\biggl(1 - |\beta - \frac{1}{2}|\biggr)$ & $\frac{\delta(1 + |\beta - \frac{1}{2}|)}{2|\beta - \frac{1}{2}|}$ \\ 
\hline
$\text{ord}(1), |\beta-1/2|\gg \delta $ & $1 - \sqrt{\frac{\delta}{2\alpha_1(\beta - \frac{1}{2})}}$ & $ \frac{1}{2}\biggl(1 + \sqrt{\frac{\delta}{2\alpha_1(\beta - \frac{1}{2})}}(\alpha_1(\beta - \frac{1}{2}) - 1)\biggr)$ & $\sqrt{\frac{\delta \alpha_1(\beta - \frac{1}{2})}{2}}$ \\ 
\hline
$\text{ord}(1), \beta=1/2 $ & $1 - \biggl( \frac{\delta}{2\alpha_1}\biggr)^{1/3}$ & $\frac{1}{2}\biggl(1 -  \biggl( \frac{\delta}{2\alpha_1}\biggr)^{1/3}\biggr)$ &  $\biggl(\frac{\delta^2\alpha_1}{4}\biggr)^{1/3}$\\ 
\hline
$\text{ord}(1/\epsilon)$ & $1$ & $\frac{1}{2}\biggl(1 + \sqrt{\frac{\delta\alpha_1 \beta}{2}}\biggr)$ & $\sqrt{\frac{\delta\alpha_1 \beta}{2}}$ \\
\hline
$\text{ord}(1/\epsilon^2)$& $1$ & $1 -\frac{\alpha_2}{4\alpha_1 \beta}$ & $1 -\frac{\alpha_2}{2\alpha_1 \beta}$ \\
\botrule
\end{tabular}
\end{table}\par
The expressions in Table {\ref{Summary table}} present relationships between model parameters and the quantities of interest depending on the magnitude of $\beta$. These asymptotic approximations can be used for parameter estimation by fitting the above expressions to data and to give intuition on how changing model parameters can impact quantities of interest. In particular, the expressions in Table {\ref{Summary table}} show that in all cases $A_1 + 2A_2 \approx 2W$ and the number of monovalently bound antibodies is small, unless $\beta=\text{ord}(\epsilon^{-2})$. This suggests that the only way to enhance monovalent binding and increase the number of bound antibodies, is to have a high antibody concentration. However, large antibody concentrations may result in higher toxicity for the patient. Therefore a balance needs to be found between maximising bound antibody number and ensuring toxicity is tolerable. \par
It is worth noting that the asymptotic analysis shows that model predictions are classified in generality despite the large parameter space. In particular, we have shown there is a simple mechanism that underlies the complex binding behaviours observed. If parameters such as the target antigen density, $r_{\text{tot}}$, were to change, the underlying binding mechanisms would not differ though this would have an impact on the antibody concentration, $A_{\text{init}}$, required to produce the same magnitude of $\beta$. \par
When $\beta = \text{ord}(1)$ and $\beta = \text{ord}(\epsilon^{-1})$,  we are able to assume the number of bivalently bound antibodies, $A=A_2$, quickly reaches quasi-static equilibrium, to obtain leading order approximations to the solutions of the model equations. In contrast, for $\beta = \text{ord}(\epsilon^{-2})$ both variables $W$ and $A$ initially change in the same timescale and hence represents a more complex case. In fact, if we considered $\beta = \text{ord}(\epsilon^{-3})$ then we could again assume one of the variables is quasi-steady at the earliest times and simplify the analysis. Overall, the results of our asymptotic analysis were in good agreement with the numerical simulations. \par
A potential limitation of the model is that we have neglected antigen internalisation. This process typically happens on a longer timescale so may affect the long time values presented in Table {\ref{Summary table}}. Since in vitro antibody binding experiments typically run over a few hours the impact of internalisation may be small compared to the in vivo setting \mbox{\citep{Birtwistle2009, Vainshtein2014, Mazor2016}}.\par
To summarise, we have presented an analysis of a model of bivalent antibody-antigen binding. Using asymptotic analysis, we have described the complex binding dynamics of bivalent antibody-antigen binding for a wide range of antibody concentrations. For the case of antigen being in excess of antibody, we provide asymptotic approximations to the model simulations that are in good agreement with the numerics. Alternatively, when antibody is in excess of antigen, our analysis shows that, within the outer region, the antibody-antigen interactions contain two separate regions (termed Regions I and II) when antibody is in abundance of antigen. Regions I and II are characterised by an abundance or lack of availability of unbound antigen respectively. With the transition from Region I to Region II, the dominant balance within the model equations change and an antibody can dissociate one of its binding arms. We use the results of the asymptotic analysis to derive simple expressions for quantities that impact mAb treatment potency and efficacy such as antigen occupancy, of particular pertinence to immune checkpoint inhibition, and bound antibody number, which in contrast is important for ADCC.
Future work could involve extending our results to the case of bispecific antibodies and binding within the immune synapse.

\bmhead{Acknowledgements}
LAH is supported by funding from the Engineering and Physical Sciences Research Council (EPSRC) [grant number EP/S024093/1]. For the purpose of Open Access, the author has has applied a CC BY public copyright licence to any Author Accepted Manuscript (AAM) version arising from this submission.

\section*{Declarations}

\subsection*{Conflict of Interest}
The authors declare no conflict of interest.
\subsection*{Code availability}
All model simulations were conducted in Python with scripts available by contacting luke.heirene@maths.ox.ac.uk.

\begin{appendices}

\section{Demonstration That $F(\zeta)$ Has a Unique Root}\label{f proof}
Here, we show that $F(\zeta)$, defined as
\begin{equation}
    F(\zeta) = \frac{\alpha_1}{\sqrt{\zeta^2 + 2\delta} -\delta + \zeta}\biggl[ \frac{2\delta(1 + \sqrt{\zeta^2 + 2\delta} + \delta/2)}{\alpha_1} - (\sqrt{\zeta^2 + 2\delta} + \zeta)^2(\bar{\beta} + \zeta)\biggr], \label{F appendix}
\end{equation}
has a unique root $\zeta^*$ within its physiological range and, as a result, the dynamics within Region II reach a real, positive steady state. This problem is equivalent to showing that there is a unique value $\zeta^*$ such that
\begin{equation}
   \frac{2\delta}{\alpha_1} =G(\bar{\beta}, \delta, \zeta) \coloneqq \frac{(\sqrt{\zeta^2 + 2\delta} - \delta + \zeta)^2(\bar{\beta}+\zeta)}{1 + \sqrt{\zeta^2+2\delta} + \delta/2}.
\end{equation}
By showing that $2\delta/\alpha_1$ is a value attained by $G(\bar{\beta}, \delta, \zeta)$ within the physiological range of $\zeta$ and that $G(\bar{\beta}, \delta, \zeta)$ is monotonically increasing on this interval, then it follows by the intermediate value theorem that there will be a unique value $\zeta^*$ where $G(\bar{\beta}, \delta, \zeta)=2\delta/\alpha_1$.\par
First, the $\sqrt{\zeta^2 + 2\delta} - \delta + \zeta$ term in Equation (\ref{F appendix}) gives rise to a singularity when $\zeta = -1 + \delta/2$. However, we can conclude that $\zeta^*$ does not lie in the region $\zeta \in (-1, -1 + \Delta)$ and exclude it for all $\Delta \approx \mathcal{O}(\delta^{1/2})$, thereby avoiding the singularity and proceed with Equation (\ref{F appendix}). To see this, substituting $\zeta=-1+\Delta$, $\Delta \approx \mathcal{O}(\delta^{1/2})$, $\Delta>0$ into Equation (\ref{F expans}) gives
\begin{equation}
    F(\zeta)\rvert_{\zeta = -1+\Delta} = (2 + \mathcal{O}(\Delta)) - \alpha_1(\delta\Delta+ \mathcal{O}(\delta^2))(2\beta + \mathcal{O}(\Delta)). \label{F restrict range}
\end{equation}
For Equation (\ref{F restrict range}) to have a root, we require that $2\alpha_1\beta \approx \mathcal{O}(2/(\delta\Delta))$ but $\alpha_1$ and $2\beta$ are both $\text{ord}(1)$. Therefore, $F(\zeta)$ does not have a root in $(-1, -1 + \Delta)$, $\Delta \approx \mathcal{O}(\delta^{1/2})$. \par
Next, note that the physiological range for $\zeta$ is given by 
\begin{equation}
    0 \leq W = \frac{1}{2}(1-\zeta) \leq \text{min}(\beta, 1),
\end{equation}
i.e, the number of bound antibodies, $W$, is bounded by either the available number of antibodies or antigens. From the above inequality,  when $\beta<1$:
\begin{equation}
    \bar{\beta} \leq \zeta \leq 1,
\end{equation}
where $\bar{\beta} = \beta - 1/2$, and when $\beta \geq 1$:
\begin{equation}
    \text{max}(-1, -\bar{\beta}) \leq \zeta \leq 1.
\end{equation}
In order to avoid the previously described singularity which is also in an unphysiological range, this inequality becomes
\begin{equation}
    \text{max}(-1 + \delta^{1/2}, -\bar{\beta}) \leq \zeta \leq 1.
\end{equation}
We also note that  $\zeta^* > -\bar{\beta}$ as $G(\bar{\beta}, \delta, \zeta)\leq 0 $ for $\zeta \leq -\bar{\beta}$ and we require that $G(\bar{\beta}, \delta, \zeta)> 0$ for a root.\par
It follows that when $\beta<1$:
\begin{align}
    &G(\bar{\beta}, \delta, -\bar{\beta}) = 0 < \frac{2\delta}{\alpha_1}, \\
    &G(\bar{\beta}, \delta, 1) = \frac{(2 + \mathcal{O}(\delta))(\beta + 1/2)}{2(2 + \mathcal{O}(\delta)} > \frac{2\delta}{\alpha_1}.
\end{align}
Similarly, when $\beta \geq 1$, $\bar{\beta}\geq 1/2$ and $-\bar{\beta}< -1 + \delta^{1/2}$ (the analysis is analogous for  $-\bar{\beta} > -1 + \delta^{1/2}$),
\begin{align}
    &G(\bar{\beta}, \delta, -1 + \delta^{1/2}) = \frac{(\delta^{3/2}(1 + o(1)))^2(\bar{\beta} + \delta^{1/2})}{1+\mathcal{O}(\delta)} \approx \mathcal{O}(\delta^3) < \frac{2\delta}{\alpha_1}, \\
   &G(\bar{\beta}, \delta, 1) = \frac{(2 + \mathcal{O}(\delta))(\beta + 1/2)}{2(2 + \mathcal{O}(\delta)} > \frac{2\delta}{\alpha_1}.
\end{align}
Having showed that $G$ attains values less and more than $2\delta/\alpha_1$ at the ends of the physiological ranges of $\zeta$, It only remains to show that $G$ is monotonic within these ranges. Before doing so, for later use, we estimate the sign of $\sqrt{\zeta^2 + 2\delta} - \delta + \zeta$:
\begin{equation}
    (\sqrt{\zeta^2 + 2\delta} - \delta + \zeta)\rvert_{\zeta=-1 + \delta^{1/2}}= \delta^{3/2}(1+o(1))>0, \label{zeta eval}
\end{equation}
and it can also be shown that the derivative of Equation (\ref{zeta eval}) with respect to $\zeta$  is strictly positive. Therefore, $\sqrt{\zeta^2 + 2\delta} - \delta + \zeta$ is always positive for the range of $\zeta$ we consider here. \par
We now consider the sign of $\mathrm{d}G/\mathrm{d}\zeta$:
\begin{equation}
\begin{split}
    &\frac{\mathrm{d}G}{\mathrm{d}\zeta} = \frac{(\sqrt{\zeta^2 + 2\delta} - \delta + \zeta)^2}{1 + \delta/2 + \sqrt{\zeta^2 + 2\delta}} + 
    \frac{(\bar{\beta}+\zeta)(\sqrt{\zeta^2 + 2\delta} - \delta + \zeta)}{(1 + \delta/2 + \sqrt{\zeta^2 + 2\delta})^2(\sqrt{\zeta^2 + 2\delta})} \times \\ &\biggl[\underbrace{2(1 + \delta/2 + \sqrt{\zeta^2 + 2\delta})(\zeta + \sqrt{\zeta^2 + 2\delta}) - \zeta(\sqrt{\zeta^2 + 2\delta} - \delta + \zeta)}_{\mathclap{\eqqcolon H}}\biggr]. \label{G deriv}
\end{split}
\end{equation}
The term multiplying $H$ is positive. It only remains to show that $H$ is positive within the required range of $\zeta$ to prove that $G$ is monotonically increasing. Analysing the expression for $H$:
\begin{align}
    H &=2(1 + \delta/2 + \sqrt{\zeta^2 + 2\delta})(\zeta + \sqrt{\zeta^2 + 2\delta}) - \zeta(\sqrt{\zeta^2 + 2\delta} - \delta + \zeta), \\
    \begin{split}
         &= \zeta(\zeta + \sqrt{\zeta^2 + 2\delta})  - \zeta(\sqrt{\zeta^2 + 2\delta} - \delta + \zeta) +  \\
         &2(1 + \delta/2 + \sqrt{\zeta^2 + 2\delta})(\zeta + \sqrt{\zeta^2 + 2\delta}) - \zeta(\zeta + \sqrt{\zeta^2 + 2\delta}), \label{H step 1}
    \end{split}
\end{align}
where we have added and subtracted  $\zeta(\zeta + \sqrt{\zeta^2 + 2\delta})$. Grouping together the first and second term in Equation (\ref{H step 1}) gives
\begin{align}
    H &= \delta\zeta + (\zeta + \sqrt{\zeta^2 + 2\delta})[2 + \delta + 2\sqrt{\zeta^2 + 2\delta} - \zeta], \\
    &> \delta\zeta +  (\zeta + \sqrt{\zeta^2 + 2\delta})(2 + \delta + \sqrt{\zeta^2 + 2\delta}). \label{H step 2}
\end{align}
To show that $H$ is positive, we need to show that the second term in Equation (\ref{H step 2}) is always larger than $\delta \zeta$ across the range of $\zeta$ we are interested in. Noting that $\zeta \geq -1 + \delta^{1/2}$, consider:
\begin{equation}
    \frac{\mathrm{d}}{\mathrm{d}\zeta}(\zeta + \sqrt{\zeta^2 + 2\delta}) = \frac{1}{\sqrt{\zeta^2 + 2\delta}} \biggl[\underbrace{\sqrt{\zeta^2 + 2\delta}}_{\mathclap{>|\zeta|}} + \zeta\biggr] >0,
\end{equation}
so $\zeta + \sqrt{\zeta^2 + 2\delta}$ is an increasing function of $\zeta$. If we can now show that at $\zeta = -1 + \delta^{1/2}$, the bottom end of the range of $\zeta$ we are considering, that $\zeta + \sqrt{\zeta^2 + 2\delta} > \delta$, then it follows from the fact that $2 + \delta + \sqrt{\zeta^2 + 2\delta} > \zeta $ that $H$ is positive. Therefore, letting $\zeta=-1 + \delta^{1/2} = -1 + \Delta$ we have
\begin{align}
    \zeta + \sqrt{\zeta^2 + 2\delta} &=  -1 + \Delta + |-1+\Delta|\sqrt{1 + \frac{2\delta}{(1-\Delta)^2}}, \\
    &\approx (-1+\Delta) + (1-\Delta) + \frac{\delta}{1-\Delta} - \frac{\delta^2}{(1-\Delta^3)} + o(\delta^3),\\
    &= \delta(1+\Delta + \Delta^2 + o(\Delta^3)) - \delta^2(1+ 3\Delta + o(\Delta^2)) + o(\delta^3), \\
    &= \delta + \delta^{3/2} + o(\delta\Delta^3, \Delta\delta^2, \delta^3),\\
    &= \delta + \delta^{3/2} + o(\delta^{5/2}) > \delta.
\end{align}
Therefore, $H>0$ and it follows that $\mathrm{d}G/\mathrm{d}\zeta>0$. To summarise, we have shown that $G$ is monotonically increasing on the physiological range of $\zeta$ and that the value $G$ attains at the bottom of this range is less than $2\delta/\alpha_1$ and vice versa for the top of the range. It follows by the intermediate value theorem that there exists a unique value of $\zeta=\zeta^*$ such that $G(\bar{\beta}, \delta, \zeta^*)=2\delta/\alpha_1$ and $\zeta^*$ is the unique root of $F(\zeta)$. 
\section{Calculation of $\zeta^*$ for different values of $\beta$}\label{zeta appendix}
Here we give details of the calculation of the unique root of $F(\zeta)=0$, denoted $\zeta^*$, for different values of $\beta$.
\subsection{$\beta=\text{ord}(1),\beta < 1/2$, $|\beta-1/2|\gg \epsilon^2=\text{ord}(\delta)$}\label{sec: zeta beta<1/2}
In this case the number of antigens is slightly more than the number of antibodies (i.e. $\bar{\beta} < 0$). With $\beta<1/2$, antigen levels exceed antibody levels. Before the antibodies are close to running out, $|W-\beta|\gg \epsilon^2$, the dynamics of $A$ and $W$ are governed by Equation (\ref{beta neq 1/2 sol}) and $A_1=0$ at leading order. To obtain $\zeta^*$ for this value of $\beta$, we will use the formula for the gradient of a line at the point $\zeta = -\bar{\beta}$. Calculating the gradient of $G$ gives
\begin{equation}
    \frac{dG}{d\zeta}\biggr\rvert_{\zeta=-\bar{\beta}}=\frac{\biggl(\sqrt{|\bar{\beta|}^2 + 2\delta} -\delta + |\bar{\beta}|\biggr)^2}{1 + \sqrt{|\bar{\beta|}^2 + 2\delta} + \delta/2} \approx \frac{4 |\bar{\beta}|^2}{1 + |\bar{\beta}|},
\end{equation}
where we have used the fact that $|\bar{\beta}| \gg \delta^{1/2}$. It follows that
\begin{equation}
    \frac{\Delta G}{\Delta \zeta} = \frac{2\delta/\alpha_1}{\zeta^* - |\bar{\beta}|} \approx \frac{4 |\bar{\beta}|^2}{1 + |\bar{\beta}|}, \label{grad eqn1}
\end{equation}
where $\Delta G = G(\zeta=-\bar{\beta}) - G(\zeta=\zeta^*)$. From Equation (\ref{grad eqn1}) it follows that the root of $F(\zeta)=0$, $\zeta^*$, is
\begin{equation}
    \zeta^* \approx |\bar{\beta}| + \frac{\delta(1 + |\bar{\beta}|)}{2|\bar{\beta}|^2\alpha_1} + \mathcal{O}(\delta^2),
\end{equation}
so that
\begin{equation}
    W = \frac{1}{2}\biggl(1 - \zeta\biggr) \rightarrow  \frac{1}{2}\biggl(1 - |\bar{\beta}|\biggr) \text{   as   } \tau \rightarrow \infty.
\end{equation}
From Equation (\ref{A1 region 2}), noting $\bar{\beta} <0$, we have that
\begin{equation}
    A_1 \rightarrow \frac{\delta(1+|\bar{\beta}|)}{2|\bar{\beta|}} \text{   as   } \tau \rightarrow \infty.
\end{equation}
Antigen occupancy for this value of $\beta$ is estimated to be
\begin{align}
    &A_1 + 2A_2 \approx 2A_2 , \\
    &= 1 - \sqrt{\zeta^2 + 2\delta} + \delta \approx 1 - |\bar{\beta}|.
\end{align}
So antigens are not fully occupied.
\subsection{$\beta=\text{ord}(1),\beta > 1/2$, $|\beta-1/2|\gg \epsilon^2=\text{ord}(\delta)$}\label{sec: zeta beta>1/2}
In this case there are slightly more antibodies than antigens. In Region I, the dynamics for $A$ and $W$ are governed by Equation (\ref{beta neq 1/2 sol}) with $A_1=0$ at leading order. For this value of $\beta$, to obtain $\zeta^*$ we make the ansatz
\begin{equation}
    \zeta^* = -\delta^{1/2}p, \label{ansatz1}
\end{equation}
with $p>0$. Substituting Equation (\ref{ansatz1}) into Equation (\ref{G eqn}) and simplifying gives
\begin{equation}
    p = \sqrt{\frac{1}{2\alpha_1\bar{\beta}}}(\alpha_1\bar{\beta} -1).
\end{equation}
It follows that within Region II
\begin{equation}
    \zeta^* = -\sqrt{\frac{\delta}{2\alpha_1\bar{\beta}}}(\alpha_1\bar{\beta} - 1),
\end{equation}
and so 
\begin{equation}
    W  \rightarrow  \frac{1}{2}\biggl(1 + \sqrt{\frac{\delta}{2\alpha_1\bar{\beta}}}(\alpha_1\bar{\beta} - 1)\biggr),
\end{equation}
with
\begin{equation}
    A_1 \rightarrow \sqrt{\frac{\delta \alpha_1}{2}\biggl(\bar{\beta} -\sqrt{\frac{\delta}{2\alpha_1\bar{\beta}}}(\alpha_1\bar{\beta} - 1)\biggr)} \approx \sqrt{\frac{\delta \alpha_1 \bar{\beta}}{2}} \text{   as   } \tau \rightarrow \infty.
\end{equation}
We note also from Figure \ref{fig: G figs} that $\zeta^* <0$ as $\alpha_1\bar{\beta}>1$. Antigen occupancy is then
\begin{align}
   &A_1 + 2A_2 = 2W - A_1, \notag \\
   &\approx 1 + \sqrt{\frac{\delta}{2\alpha_1\bar{\beta}}}(\alpha_1\bar{\beta} - 1) - \sqrt{\frac{\delta \alpha_1 \bar{\beta}}{2}}, \notag \\
   &= 1 - \sqrt{\frac{\delta}{2\alpha_1 \bar{\beta}}}.
\end{align}
So antigens are fully occupied at leading order.
\subsection{$\beta=\text{ord}(1),\beta = 1/2$}\label{sec: zeta beta=1/2}
In this special case the number of antigens and antibody are equal. Although this case may be physically unrealistic due to the need for extreme parameter fine-tuning, we include it for completeness. In Region I, the dynamics for $A$ and $W$ are governed by Equation (\ref{beta=1/2 sol}) with $A_1=0$ at leading order.  For this value of $\beta$, to obtain $\zeta^*$ we make the ansatz
\begin{equation}
    \zeta^*=\delta^ap, \label{ansatz}
\end{equation}
with $p>0$. Substituting Equation (\ref{ansatz}) into Equation (\ref{G eqn}) and simplifying gives
\begin{equation}
    4\delta^{2a}p^2 + 4\delta + \frac{\delta^{2-2a}}{p^2} = \frac{2\delta^{1-a}}{\alpha_1p}.
\end{equation}
Finding the dominant balance in the above equation gives $a=1/3$. It then follows that
\begin{equation}
   \frac{2\delta^{2/3}}{\alpha_1p} = 4\delta^{2/3}p^2 + \text{h.o.t} \implies p \approx \biggl(\frac{1}{2\alpha_1}\biggr)^{1/3}.
\end{equation}
Therefore, as antigens run out, we have
\begin{equation}
    \zeta^* \approx \biggl( \frac{\delta}{2\alpha_1}\biggr)^{1/3},
\end{equation}
so that
\begin{equation}
    W  \rightarrow  \frac{1}{2}\biggl(1 -  \biggl( \frac{\delta}{2\alpha_1}\biggr)^{1/3}\biggr) \text{   as   } \tau \rightarrow \infty,
\end{equation}
while Equation (\ref{A1 region 2}) gives
\begin{equation}
    A_1 \rightarrow \biggl(\frac{\delta^2\alpha_1}{4}\biggr)^{1/3} \text{   as   } \tau \rightarrow \infty.
\end{equation}
Antigen occupancy is then given by
\begin{align}
     &A_1+ 2A_2 \approx 2W, \\
     &= 1-\biggl(\frac{\delta}{2\alpha_1}\biggr)^{1/3} \approx 1.
\end{align}
So once again antigens are fully occupied.
\section{Region I limits of $\lambda$ and $\zeta$ for $\beta=\text{ord}(1/\epsilon^2)$}\label{region I beta big}
\begin{figure}
    \centering
    \includegraphics[width=0.8 \textwidth]{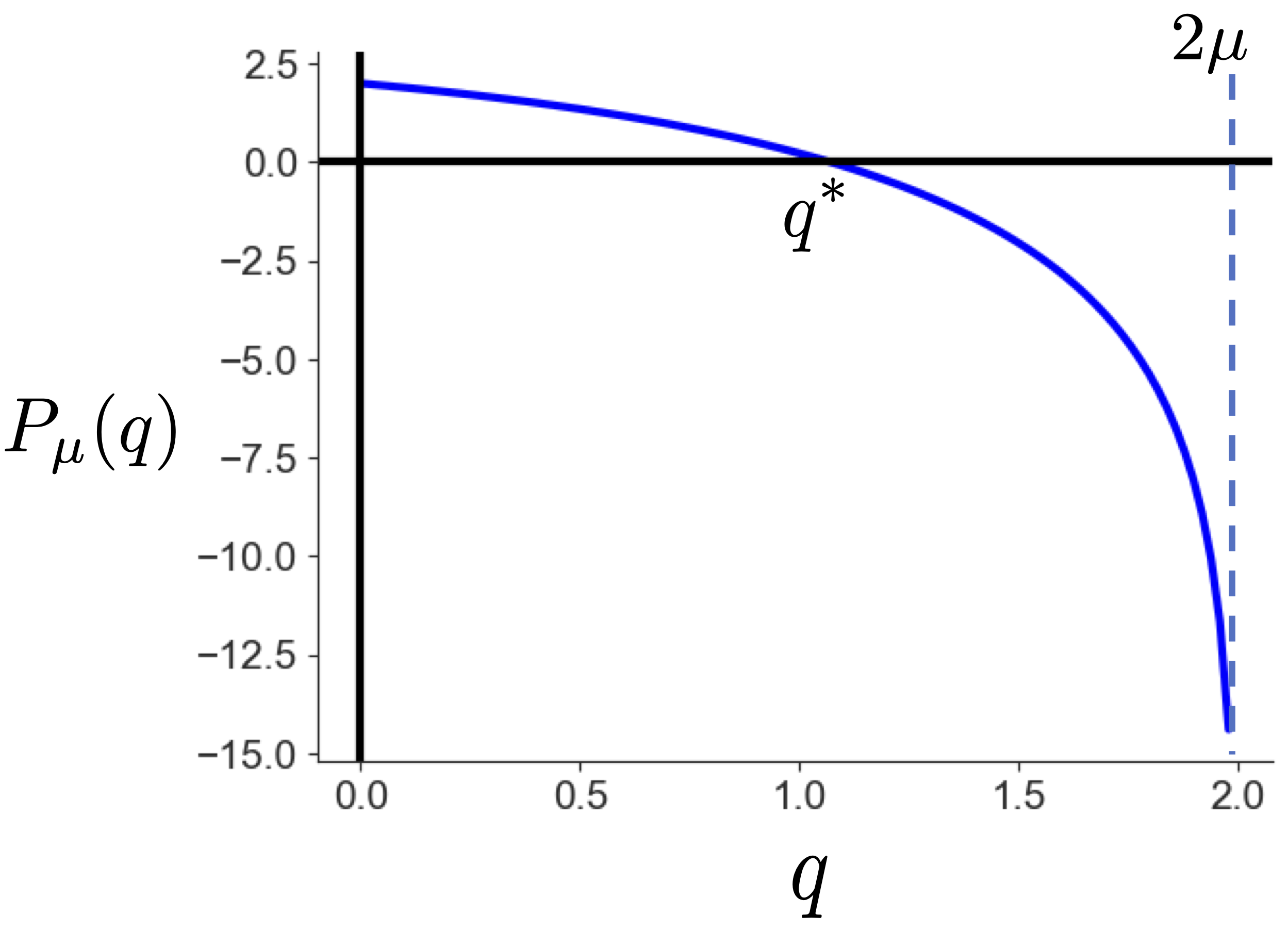}
    \caption{Plot of $P_{\mu}(q)$ given by Equation (\ref{p express}) as $q$ varies. The unique root of $P_{\mu}(q)$ is denoted $q^*$.}
    \label{fig: P(q)}
\end{figure}

For $\beta=\text{ord}(1/\epsilon^2)$, the model equations can be cast in terms of $p=\zeta+\lambda=2(1-W-A)$ and $q=\lambda-\zeta=2(W-A)$:
\begin{align}
    \frac{\mathrm{d}p}{\mathrm{d}\bar{\tau}} &= p\biggl(-\mu - \frac{q}{2}\biggr), \label{p eqn1} \\
    \frac{\mathrm{d}q}{\mathrm{d}\bar{\tau}} &= p\biggl(\mu - \frac{q}{2}\biggr), \label{q eqn1}
\end{align}
Dividing Equation (\ref{p eqn1}) by Equation (\ref{q eqn1}) and integrating we deduce
\begin{equation}
    p = 2 + q + 4\mu\text{ln}\biggl(1 - \frac{q}{2\mu}\biggr) \eqqcolon P_{\mu}(q). \label{p express}
\end{equation}
This expression is valid at small times where $q$ is small such that $q/2\mu <1$. Substituting for $p$ in Equation (\ref{q eqn1}) gives
\begin{equation}
    \frac{\mathrm{d}q}{\mathrm{d}\bar{\tau}} = \biggl(\mu-\frac{q}{2}\biggr)\biggl(2 + q + 4\mu\text{ln}\biggl(1 - \frac{q}{2\mu}\biggr)\biggr) \coloneqq F_{\mu}(q),
\end{equation}
with $q(0)=0$. We are interested in the behaviour of $F_{\mu}(q)=(\mu - q/2)P_{\mu}(q)=0$ to determine how $q$ behaves as $\bar{\tau}\rightarrow \infty$. Let us first look at a graph of $P_{\mu}(q)$ in Figure \ref{fig: P(q)}.
We can see that $P_{\mu}(0)$ = 2 and $P_{\mu}(q) \rightarrow - \infty$ as $q \rightarrow 2\mu$. Observing Equation (\ref{p express}) we see that
\begin{equation}
    \frac{\mathrm{d}P_{\mu}}{\mathrm{d}q} = 1 + \frac{4\mu}{q-2\mu} \leq -1,
\end{equation}
when $q \in (0, 2\mu)$. Therefore, $P_{\mu}(q)$ and, as a result, $F_{\mu}(q)$ has a unique root $q^*\in (0, 2\mu)$ with $q\rightarrow q^*$ as $\bar{\tau} \rightarrow \infty$. Therefore, $P_{\mu}(q) \rightarrow 0$ as $\bar{\tau} \rightarrow \infty$. But $p = P_{\mu}(q)$ so therefore $p \rightarrow 0 $ as $\bar{\tau} \rightarrow \infty$. From this
\begin{equation}
    p \rightarrow 0 \text{   as   } \tau \rightarrow \infty. \label{inner asymptote}
\end{equation}
Finally, as a result we have
\begin{equation}
    A+W \rightarrow 1 \text{   as   } \tau \rightarrow \infty,
\end{equation}
\section{Region II limits of $\lambda$ and $\zeta$ for $\beta=\text{ord}(1/\epsilon^2)$}\label{region II beta big}
Rewriting Equations (\ref{W 1/epsilon^2}) and (\ref{A 1/epsilon^2}) in terms of $\zeta$ and $\lambda$ we have 
\begin{align}
    \frac{\mathrm{d}\zeta}{\mathrm{d} \tau} &= -\frac{\mu}{\epsilon^2}(\zeta + \lambda) + \frac{(\lambda - \zeta)}{\hat{\alpha_2}}, \label{zeta full}\\
    \frac{\mathrm{d} \lambda}{\mathrm{d} \tau} &= -\frac{1}{2\epsilon^2}(\lambda-\zeta)(\zeta+\lambda) + \frac{2}{\hat{\alpha_2}}(1-\lambda). \label{lambda full}
\end{align}
Noting Equation $\lambda + \zeta =\mathcal{O}(\epsilon^2)$ from (\ref{zeta lambda approx}), we define $\kappa$ via
\begin{equation}
    \lambda = -\zeta  + \epsilon^2\kappa. \label{lambda kappa relations}
\end{equation}
Rewriting Equations (\ref{zeta full}) and (\ref{lambda full}) in terms of $\zeta$ and $\kappa = (\lambda + \zeta)/\epsilon^2$ we arrive at
\begin{align}
    \frac{\mathrm{d}\zeta}{\mathrm{d}\tau} &= -\mu \kappa - \frac{2\zeta}{\hat{\alpha_2}} + \mathcal{O}(\epsilon^2), \label{zeta kappa eqn1}\\ 
    \frac{\mathrm{d}\kappa}{\mathrm{d}\tau} &= \frac{1}{\epsilon^2}\biggl[(\zeta-\mu)\kappa + \frac{2}{\hat{\alpha_2}} + \mathcal{O}(\epsilon^2) \biggr], \label{kappa ode1}
\end{align}
where $\mu=2\alpha_1 \hat{\beta}/\hat{\alpha_2}=\text{ord}(1)$ and positive. Further, noting Equations ({\ref{zeta lambda approx}}) and ({\ref{lambda kappa relations}}) we have $\lambda > \zeta$ and $\lambda  =\kappa \epsilon^2 - \zeta$ with $\kappa = \mathcal{O}(1)$.  It follows that $\zeta < \kappa \epsilon^2$ and therefore $\zeta \ll \mu = \text{ord}(1)$. \par
From Equations (\ref{zeta kappa eqn1}) and (\ref{kappa ode1}), we see that $\kappa$ evolves on a much faster timescale than $\zeta$. As such, $\kappa$ will reach its steady state, $\kappa^*$, which at leading order is given by
\begin{equation}
    \kappa^* = \frac{2}{\hat{\alpha}_2}\biggl(\frac{1}{\mu-\zeta}\biggr) \approx \frac{2}{\hat{\alpha}_2 \mu}. \label{kappa stst}
\end{equation}
Note that $\kappa$ reaches $\kappa^*$ from $\kappa(0)=0$ after the system enters the outer region with $\kappa(0)=0$, representing the standard van Dyke condition of matched asymptotic expansions \mbox{\citep{Hinch1991}}, obtained by matching the long time asymptote of the inner solution that is given by Equation ({\ref{inner asymptote}}).
Substituting Equation (\ref{kappa stst}) for $\kappa = \kappa^*$ in Equation (\ref{zeta kappa eqn1}) gives at leading order
\begin{equation}
    \frac{\mathrm{d}\zeta}{\mathrm{d}\tau} = -\frac{2}{\hat{\alpha_2}}\biggl(\frac{\mu}{\mu - \zeta} + \zeta \biggr).\label{zeta eqn final}
\end{equation}
Setting $\mathrm{d}\zeta/\mathrm{d}\tau =0$ we have after simplifying
\begin{equation}
    \zeta^2 - \mu\zeta - \mu = 0,
\end{equation}
which has roots 
\begin{equation}
    \zeta_{\pm} = \frac{\mu}{2}\biggl(1 \pm \sqrt{1 + \frac{4}{\mu}}\biggr). \label{zeta sol}
\end{equation}
From previous analysis we know that $\zeta<\mu=\text{ord}(1)$. Thus by plotting the function $K(\zeta)=-(\mu/(\mu-\zeta) + \zeta)$ one can deduce that $\zeta \rightarrow \zeta_-$ at large time. Substituting $\zeta=\zeta_-$ and $\kappa=\kappa^*$ into Equation (\ref{lambda kappa relations}) we arrive at
\begin{equation}
    \lambda \approx - \frac{\mu}{2}\biggl(1 - \sqrt{1 + \frac{4}{\mu}}\biggr) + \epsilon^2\frac{2}{\hat{\alpha}_2 \mu} \text{   as   } \tau \rightarrow \infty. \label{lambda sol}
\end{equation}




\end{appendices}


\bibliography{Asymptotics}

\end{document}